\documentclass[journal]{IEEEtran}

\usepackage{amsmath}
\usepackage{tabularx}
\usepackage{graphicx}
\usepackage{subfigure}
\usepackage{amssymb}
\usepackage{amsthm}
\usepackage{color}
\usepackage{algorithm,algpseudocode}
\newtheorem{theorem}{Theorem}
\newtheorem{corollary}{Corollary}
\newtheorem{proposition}{Proposition}
\newtheorem{lemma}{Lemma}
\newtheorem{definition}{Definition}

\usepackage{url} 
\usepackage{color}
\usepackage{cite}

\begin{document}

\title{Optimal Curing Strategy for Competing Epidemics Spreading over Complex Networks}

\author{ Juntao Chen,~\IEEEmembership{Member,~IEEE}, Yunhan Huang,~\IEEEmembership{Student Member,~IEEE}, Rui Zhang, and
          Quanyan~Zhu,~\IEEEmembership{Member,~IEEE}

\thanks{This paper has been accepted for publication in \textit{IEEE Transactions on Signal and Information Processing over Networks}.}
\thanks{This work was supported in part by the National Science of Foundation under Grant ECCS-1847056, Grant CNS-1544782, Grant CNS-2027884, and Grant SES-1541164. The work of J. Chen was partially supported by a Faculty Research Grant from Fordham Office of Research.}          
\thanks{J. Chen is with the Department of Computer and Information Sciences, Fordham University,  New York,  NY 10023 USA, E-mail: jchen504@fordham.edu.}
\thanks{Y. Huang, R. Zhang, and Q. Zhu are with the Department of Electrical and Computer Engineering, Tandon School of Engineering, New York University, Brooklyn, NY 11201 USA. E-mail: \{yh2315, rz885, qz494\}@nyu.edu.}}% <-this % stops a space

\maketitle

\begin{abstract}
Optimal curing strategy of suppressing competing epidemics spreading over complex networks is a critical issue. In this paper, we first establish a framework to capture the coupling between two epidemics, and then analyze the system's equilibrium states by categorizing them into three classes, and deriving their stability conditions. The designed curing strategy globally optimizes the trade-off between the curing cost and the severity of epidemics in the network. In addition, we provide structural results on the predictability of epidemic spreading by showing the existence and uniqueness of the solution. We also demonstrate the robustness of curing strategy by showing the continuity of epidemic severity with respect to the applied curing effort. A gradient descent algorithm based on a fixed-point iterative scheme is proposed to find the optimal curing strategy. Depending on the system parameters, the curing strategy can lead to switching between equilibria of the epidemic network as the control cost varies. Finally, we use case studies to corroborate and illustrate the obtained theoretical results.
\end{abstract}

\begin{IEEEkeywords}
Optimal curing, Epidemic spreading, Complex networks, Competing epidemics
\end{IEEEkeywords}
%\end{frontmatter}

\section{Introduction}
With the growth of urban population and the advances in technologies and infrastructures, our world becomes highly connected, and witnesses fast economic development. The connectivity not only enables the communications among mobile and networked devices but also creates dense social and physical interactions in societies, resulting in densely connected complex networks. 
As the connectivity facilitates the information exchange and the social interactions, its also allows diseases and viruses to spread over the network in multifarious ways. For example, the WannaCry Ransomware has spread through the Internet and infected more than 230,000 computers in over 150 countries. The spreading of Ebola disease in 2014 from West Africa to other countries such as US, UK, and Spain relies on the global connectivity. The undergoing COVID-19 pandemic around the world also reflects the vulnerability of connectivity.

Control of the epidemics of diseases and computer viruses is an essential way to mitigate their social and economic impact. Depending on the nature of the epidemics, we can design centralized or distributed policies to contain the growth of the infected population by protecting, removing, and recovering nodes from the population. In human networks where HIN1, HIV, and Ebola viruses can spread, vaccine allocations will be an effective control mechanism. In computer networks that are vulnerable to malware, anti-virus software and quarantine strategies play an essential role in assuring network security.

The control of homogeneous epidemics has found applications in viral marketing \cite{moreno2004dynamics}, computer security \cite{omic2009protecting,garetto2003modeling}, and epidemiology \cite{pastor2001epidemic,gross2006epidemic,preciado2013optimal}. Note that in the control of epidemics, the underlying network structure plays an essential role as it provides communication and networking between nodes which is reflected in application scenarios such as computer and social networks \cite{pastor2002immunization,ganesh2005effect,zhang2007performance}. With the integration of multiple technologies and the growing complexity of the network systems, homogeneous epidemic models are not sufficient to capture the coexistence of heterogeneous epidemic processes. For example, it has been shown that influenza viruses can mutate and circulate in the human population during the epidemic season. An individual is not likely being infected by multiple strains simultaneously. Once an individual is infected by one type, he cannot be infected by a virus of a different type \cite{karrer2011competing}. Similarly, in the marketing over social media, two similar products will compete for their customers by spreading information over social networks. An individual who has bought one kind of product is not likely to purchase the same product from another manufacturer.  Therefore, it is essential to address the heterogeneous control of interdependent epidemics in a holistic framework.
 
To this end, this work focuses on the optimal control\footnote{The control effort refers to the applied curing strategy which we use interchangeably in the paper.} of two interdependent epidemics with a competing mechanism spreading over complex networks. To capture the dynamics of the epidemics, we use an susceptible-infected-susceptible (SIS) epidemic model for both epidemic processes of two strains of viruses 1 and 2, which leads to an epidemic model with three states: (i) susceptible (healthy), (ii) infected by strain 1, and (iii) infected by strain 2. Those infected entities can be treated and moved to the susceptible state through control. 
 
We first study the steady state of the proposed epidemics over complex networks. Through analyzing the non-linear differential equations that model the competing epidemics with control, we observe a non-coexistence phenomenon. Specifically, the network can be in the following three possible equilibrium states: (i) only strain 1, (ii) only strain 2 and (iii) disease-free. Therefore, a coexistence of two competing epidemics in the same network is impossible at the steady state. Furthermore, we investigate the stability of each network equilibrium via the eigenvalue analysis of its linearized dynamic systems.

To design the optimal control strategy, we formulate an optimization problem that minimizes the control cost as well as the severity of epidemics over the network jointly. We propose a gradient-decent algorithm based on a fixed-point iterative scheme to compute the optimal solution and show its convergence to the corresponding fixed-point. In the disease-free regime, we provide a closed-form solution for the optimal control. One critical feature of the policy in this regime is that it is fully determined by the average degree of the epidemic network and the second moment of the degree distribution which yields a distribution independent optimal curing strategy.
We further observe one emerging phenomenon that, under some conditions, the network encounters a switching of equilibrium through optimal control as the unit cost of effort changes. Depending on the system parameters, the network can be directly controlled to the disease-free equilibrium or from one exclusive equilibrium to the other one first with the symmetric control efforts of two competing epidemics. As long as the applied effort drives the epidemic network to the disease-free equilibrium, the control effort ceases to increase though the unit control cost continues to decrease. The control effort under  which the network switches from exclusive equilibrium of strain 1 or 2 to disease-free regime is referred as \textit{fulfilling threshold}. Finally, we use several numerical experiments on a scale-free network  to corroborate the derived theoretical results and discovered phenomenon. The proposed framework and approach in this paper can provide decision support for healthcare sectors in controlling competing epidemics spreading among a large population. The equilibrium analysis greatly facilitates an accurate prediction of system states, which leads to agile and optimal decision making on curing strategies. Our results can also help cyber operators determine policies in preventing the outbreak of viruses in large-scale computer networks, creating a highly trustworthy cyberspace.

The contributions of this paper are summarized as follows.

\begin{enumerate}
\item We analyze the SIS epidemic model with two competing epidemic processes of two viruses on complex networks. We characterize the equilibrium solution of the epidemics and observe a non-coexistence phenomenon, i.e., only one of the viruses or none of them will exist at the steady state.
\item We derive the necessary and sufficient conditions for the stability of three types of network equilibria: the exclusive equilibria of two viruses and the disease-free equilibrium.
\item We formulate an optimization problem to find the optimal control strategies for different regimes of the epidemic process and propose an iterative algorithm to compute the optimal curing strategy.
\item We show the predictability of epidemic spreading by proving the existence and uniqueness of epidemic level solution to the formulated problem. Furthermore, we demonstrate the robustness of curing strategy by showing the continuity of epidemic severity with respect to the applied effort.
\item We study the switching phenomena between network equilibrium under optimal curing strategy when the unit control cost changes and present explicit conditions for single and double switching of the equilibria under a class of symmetric control.
\item We formally define fulfilling threshold which refers to the optimal control strategy that drives the epidemic network to the stable disease-free equilibrium. We explicitly identify the fulfilling threshold which serves as an upper bound on the network operator's control effort.
\end{enumerate}

\subsection{Related Work}

The growing social and computer networks provide a fertile medium for the spreading of epidemics. A number of previous works have been on modeling the dynamic processes of epidemics including \cite{pastor2001epidemic,van2009virus,gang2005epidemic,newman2002spread}.  More recently, a growing number of works have investigated the epidemics spreading on multiplex/interconnected networks \cite{dickison2012epidemics,saumell2012epidemic,watkins2018optimal}, and time-varying underlying epidemic networks \cite{pare2018epidemic,prakash2010virus}.  Optimal control of single strain epidemics spreading over networks has been considered in various applications including biological disease and virus  \cite{preciado2013optimal,hansen2011optimal,sahneh2011epidemic,ramirez2014distributed,nowzari2017optimal,befekadu2019optimal,huang2019differential},  and network security \cite{omic2009protecting,trajanovski2015decentralized,preciado2014optimal,hayel2017epidemic,huang2019achieving}. In addition to the single strain epidemics, the properties of competing epidemics or multi-strain epidemics under different models have been well studied in literature \cite{venturino2001effects,han2009epidemics,karrer2011competing,liu2019analysis}. Several methods have been proposed to control multi-strain epidemics over finite networks, including the mean-field approximation-based optimization, impulse control, and passivity-based approach \cite{watkins2018optimal,taynitskiy2017optimal,
lee2018adaptive,gubar2013optimal}. Our work contributes to the literature on epidemic control by focusing on the control of competing epidemics over complex networks. Some preliminary results of this work have been included in \cite{chen2019game}. This work can be extended further and the developed framework is applicable to address a number of cybersecurity problems, such as trust in the Internet of things (IoT) \cite{pawlick2018istrict}, cloud security  \cite{chen2017security}, and IoT risk management \cite{chen2019optimal,chen2019interdependent}.

One seminal work on the control of epidemics is \cite{khouzani2011optimal}. Our work differs from it significantly in multiple aspects. Firstly, the underlying dynamic models of epidemics are fundamentally different. Our work adopts the degree-based mean-field model to capture the epidemic spreading in a large population regime which preserves the statistics of network structure, while the focus of \cite{khouzani2011optimal} is the classical susceptible-infected-recovered (SIR) model where all nodes are assumed to be homogeneous. Secondly, we investigate the joint control of two competing strains of epidemics, while \cite{khouzani2011optimal} has focused on a single strain. Thirdly, the authors in \cite{khouzani2011optimal} have designed the optimal control policy using the Pontryagin's maximum principle. However, our work aims to obtain the optimal control of epidemics in the long run, and hence we focus more on the equilibrium and stability analysis of the steady states. Fourthly, our work also investigates the robustness of the optimal control strategy, as well as the switching behavior of network equilibrium under control, which are new to the control of interdependent/competing epidemics over complex networks.

Another critical factor that affects the epidemic spreading is human behavior, including collective awareness and social relationships. For example, \cite{funk2009spread} has investigated how people's awareness of disease reduces their susceptibility and influences the epidemic outbreak threshold. Some recent works have studied the impact of awareness on the epidemics on multiplex/multilayer networks \cite{granell2013dynamical,guo2016epidemic}.  Furthermore, \cite{funk2010modelling} has provided a summary of the behavioral responses to the spread of diseases and how they are incorporated into the epidemic model. Social awareness can be seen as an alternative form of control in our current work, and its explicit consideration is an interesting future direction.

\subsection{Organization of the Paper}
The rest of the paper is organized as follows. Section \ref{prolem_formulation} formulates the control of competing epidemics problem. Network equilibrium and stability analysis are presented in Section \ref{equi}. Section \ref{optimal_control} analyzes the problem and  develops an algorithm to compute the optimal control strategy. Section \ref{switching_section} shows the phenomenon of network equilibrium switching through optimal control. Section \ref{simulation} presents numerical experiments, and Section \ref{conslusion} concludes the paper. 

\section{Problem Formulation}\label{prolem_formulation}
In a complex network with a large number of agents, we consider the classical SIS model in which each agent can be in one of the following two states: susceptible (S) or infected (I). We further consider two strains of competing epidemics, strain 1 and strain 2, spreading over the network. Specifically, strains 1 and 2 are in a competing mechanism, i.e., each susceptible agent can either be infected by strain 1 or strain 2 by contacting with other corresponding infected individuals. Let $\zeta_1$ and $\zeta_2$ be the spreading rate of strain 1 and strain 2, respectively. In addition, the infected agents can recover to the susceptible state with rate $\gamma_1$ or $\gamma_2$ (with respect to strain 1 or strain 2). Besides the self-recovery mechanism, each infected agent can be controlled to return to the healthy state through efforts, e.g., allocating vaccines during flu outbreak season.

 To analyze the competing epidemic dynamics over complex networks, we consider a degree-based mean-field (DBMF) approximation model \cite{pastor2001epidemic,pastor2015epidemic}. Specifically, the DBMF model assumes that the nodes with the same number of degree/connectivity have an identical probability of being infected, as these nodes are regarded as statistically equivalent. One limitation of the DBMF model is that it destroys the adjacency matrix of the network which is usually adopted in the analysis of reasonable size networks. However, the mean-field approximation approach offers a substantial complexity reduction on the number of degrees of freedom, comparing with investigating the adjacency matrix of complex networks. Therefore, the DBMF model facilitates the analysis of epidemic spreading in a large population regime, such as social networks and computer networks. The DBMF model has also been corroborated to be powerful in studying the epidemic spreading in real world. For instance, after analyzing the real data reported by the Virus Bulletin from February 1996 to March 2000,  \cite{pastor2001epidemic} has shown that the scale-free property, a specific case based on the DBMF model, should be included in developing theory of epidemic spreading of computer viruses.
 
 To this end, we denote by $k$ the degree of a node, where $k\in \mathcal{K}:=\{0,1,2,...,K\}$, and $P(k)\in[0,1]$ by the probability distribution of node's degree in the network. Further, we adopt $I_{i,k}(t)\in[0,1]$ to represent the density of nodes at time $t$ with degree $k$ infected by  strain $i$, $i\in\{1,2\}$. Then, the dynamics of two competing epidemics with a control $\mathbf{u}:=(u_1,u_2)\in\mathbb{R}_+^2$ can be described by two coupled non-linear differential equations as follows:
\begin{equation}\label{strains}
\begin{split}
\frac{dI_{1,k}(t)}{dt}=&-\gamma_1 I_{1,k}(t)\\
&+\zeta_1k[1-I_{1,k}(t)-I_{2,k}(t)]\Theta_1(t)-u_1I_{1,k}(t),\\
\frac{dI_{2,k}(t)}{dt}=&-\gamma_2 I_{2,k}(t)\\
&+\zeta_2k[1-I_{1,k}(t)-I_{2,k}(t)]\Theta_2(t)-u_2I_{2,k}(t),
\end{split}
\end{equation}
where $(\gamma_1,\gamma_2)$ and $(\zeta_1,\zeta_2)$ are the recovery and spreading rates of two strains, respectively. The terms $-\gamma_1 I_{1,k}(t)$ and $-\gamma_2 I_{2,k}(t)$ indicate the proportion of affected nodes returned to the healthy state through recovery. The imposed control effort to suppress the epidemic spreading is reflected by the terms $-u_1I_{1,k}(t)$ and $-u_2I_{2,k}(t)$. Here, based on DBMF model, the agents with different degrees are controlled in the same manner reflected by $u_1$ and $u_2$. Note that it is also possible to design heterogeneous control for nodes of different degrees. However, in practice, identifying a group of nodes with the same degree and applying specific control to this particular group is challenging, knowing that the epidemic network is large-scale. Furthermore, treating the nodes in a homogeneous manner as in \eqref{strains} also preserves fairness when allocating recovery resources in combating the epidemics. In \eqref{strains}, the term $1-I_{1,k}(t)-I_{2,k}(t)$ captures the density of susceptible nodes with degree $k$. In addition, $\Theta_i(t)$ represents the probability of a given link connected to a node infected by strain $i$, and $\Theta_i\in [0,1]$, $i\in\{1,2\}$. Specifically, $\Theta_1(t)$ and $\Theta_2(t)$ admit the following expressions:
\begin{align}
\Theta_1(t)=\frac{\sum_{k'\in\mathcal{K}}k'P(k')I_{1,k'}(t)}{
\langle k \rangle},\label{theta1}\\
\Theta_2(t)=\frac{\sum_{k'\in\mathcal{K}}k'P(k')I_{2,k'}(t)}{
\langle k \rangle},\label{theta2}
\end{align}
where $\langle k \rangle:=\sum_k kP(k)$ is the average degree/connectivity of nodes in the network. The nominator $\sum_{k'}k'P(k')I_{i,k'}(t)$ stands for the average connectivity of individuals infected by strain $i$, $i\in\{1,2\}$. Note that $\sum_{k'}k'P(k')I_{i,k'}(t)\leq \langle k \rangle$. Therefore, in the class of agents with degree $k$, the epidemic spreading processes of strain $i$, $i\in\{1,2\}$, can be modeled by the term $\zeta_i k[1-I_{1,k}(t)-I_{2,k}(t)]\Theta_i(t)$ as in \eqref{strains}.

The network cost over a time period $[0,T]$ is captured by two terms: the control cost $c_1(\mathbf{u})$, and the severity of epidemics $c_2(w_1\bar{I}_1(t)+w_2\bar{I}_2(t))$, where $c_1:\mathbb{R}_+^2\rightarrow\mathbb{R}_+$, $c_2:\mathbb{R}_+^2\rightarrow\mathbb{R}_+$ $w_1$ and $w_2$ are two positive weighting constants. Note that both $c_1$ and $c_2$ are assumed to be continuously differentiable, convex, and monotonically increasing. When $\mathbf{u}=(0,0)$, we have $c_1(\mathbf{u}) = 0$. If there are no epidemics, then $c_2(0)=0$. Furthermore, $\bar{I}_1(t)$ and $\bar{I}_2(t)$ are defined as 
\begin{align}
\bar{I}_1(t):=\sum_{k\in\mathcal{K}} P(k)I_{1,k}(t),\label{I_1_def}\\
\bar{I}_2(t):=\sum_{k\in\mathcal{K}} P(k)I_{2,k}(t),
\end{align}
respectively, which can be interpreted as the severity of epidemics in the network.
The average combined cost of epidemics and control in the long run is given by
$
 \lim\limits_{T\rightarrow\infty}\ \frac{1}{T}\int_{0}^{T}c_2(w_1\bar{I}_1(t)+w_2\bar{I}_2(t)) dt +c_1(\mathbf{u}).
$
Hence, the system operator needs to determine $\mathbf{u}$ to minimize the aggregated epidemic and control costs. When $\bar{I}_1(t)$ and $\bar{I}_2(t)$ converge to a steady state as $T\rightarrow\infty$, the cost functions $c_1$ and $c_2$ admit constant values. Therefore, the problem of controlling competing epidemics can be formulated as follows:
\begin{equation*}
\begin{split}
(\mathrm{OP1}):\quad &\min_{\mathbf{u}}\ c_1(\mathbf{u})+c_2\left(w_1\bar{I}_1^*(u_1)+w_2\bar{I}_2^*(u_2)\right)\\
&\mathrm{s.t.}\quad \mathrm{system\ dynamics}\ \eqref{strains},
\end{split}
\end{equation*}
where $\bar{I}_1^*(u_1)$ and $\bar{I}_2^*(u_2)$ denote the densities of the strains at the steady state under the control $\mathbf{u}$. Note that the optimal control resulting from (OP1) is a time-invariant strategy. This modeling indicates that the system operator is concerned about the epidemic spreading in the long run and develops the optimal curing policy by predicting the network equilibrium state.
Prediction of the epidemic spreading also greatly facilitates the design of a high-confidence control. The system operator can predict which steady-state the epidemic dynamics will stabilize at under the corresponding applied control to the network. An accurate prediction also helps the operator design a proactive curing mechanism.

To address (OP1), we need to obtain $\bar{I}_1^*(u_1)$ and $\bar{I}_2^*(u_2)$. For convenience, we denote by
\begin{equation}\label{psi_def}
\psi_i:=\frac{\zeta_i}{\gamma_i+u_i},\ i=1,2,
\end{equation}
the effective spreading rate (ESR) of strains under the control.
Note that ESR $\psi_i$ quantifies the net spreading rate of strain $i$ over the network. However, the condition $\psi_i>1$, $i\in\{1,2\}$, alone cannot guarantee the outbreak of the epidemics, as analyzed in Section \ref{equi}.

At the steady state, $dI_{1,k}/dt=0$ and $dI_{2,k}/dt=0$. Then, from \eqref{strains}, we obtain
\begin{align}
I_{1,k}=\frac{\psi_1k\Theta_1}{1+\psi_1k\Theta_1+\psi_2k\Theta_2}\label{steady1},\\
I_{2,k}=\frac{\psi_2k\Theta_2}{1+\psi_1k\Theta_1+\psi_2k\Theta_2}.\label{steady2}
\end{align}

Therefore, with \eqref{steady1} and \eqref{steady2}, the optimal control problem (OP1) becomes
\begin{equation*}
\begin{split}
(\mathrm{OP2}):\quad \min_{\mathbf{u}}\quad &c_1(\mathbf{u})+c_2\Big(w_1\bar{I}_1^*(u_1)+w_2\bar{I}_2^*(u_2)\Big)\\
\mathrm{s.t.}\quad &{I}_{1,k}^*(u_1)=\frac{\psi_1k\Theta_1^*}{1+\psi_1k\Theta_1^*+\psi_2k\Theta_2^*},\ \forall k\in \mathcal{K},\\
&{I}_{2,k}^*(u_2)=\frac{\psi_2k\Theta_2^*}{1+\psi_1k\Theta_1^*+\psi_2k\Theta_2^*},\ \forall k\in \mathcal{K},\\
&\psi_i=\zeta_i/(\gamma_i+u_i),\ i=1,2,
\end{split}
\end{equation*}
where the variables with superscript $*$ denote the steady state values, i.e., $\Theta_i^*=\frac{\sum_{k'}k'P(k')I_{i,k'}^*(u_i)}{\langle k \rangle}$ and
$\bar{I}_i^*(u_i)=\sum_k P(k)I_{i,k}^*(u_i)$, $i\in\{1,2\}$.

In the suppression of diseases spreading, the control efforts are generally determined by a centralized authority. Thus,
our objective is to design a control strategy via solving (OP2) which jointly optimizes the control cost and the epidemics spreading level in the network.

The control cost function $c_1(\mathbf{u})$ which measures the cost of curing rate has been widely used in the control of epidemics literature \cite{nowzari2016analysis,chen2014optimal,forster2007optimizing}. Note that the adopted control cost function is not unique, and it can take other forms such as $c_1(u_1\bar{I}_1(t)+u_2\bar{I}_2(t))$. However, such modeling may not be appropriate if the system operator focuses on the average cost shown in (OP2), as the cost of control degenerates to zero if the epidemics die out under the applied control. Instead, the functional $c_1(u_1\bar{I}_1(t)+u_2\bar{I}_2(t))$ can be used when the system operator cares about the running cost of control and severity of epidemics over a considered time horizon. Comparing $c_1(\mathbf{u})$ with $c_1(u_1\bar{I}_1(t)+u_2\bar{I}_2(t))$, the former one may yield an over-penalization of the control when the system operator aims to achieve a disease-free network at the equilibrium. This concern can be mitigated by assigning a larger weight on $c_2$ over $c_1$ in the cost objective in (OP2) such that the system operator has a stronger desire to drive the network to a disease-free equilibrium. To explicitly address such concern, one can solve the epidemic control problem with a cost function $\int_{0}^{T} c_1(u_1\bar{I}_1(t)+u_2\bar{I}_2(t))+c_2(w_1\bar{I}_1(t)+w_2\bar{I}_2(t)) dt$, and we leave it as future work.

\section{Network Equilibrium and Stability Analysis}\label{equi}

To solve the problem (OP2), we first need to analyze the steady states of the epidemics. Substituting \eqref{steady1} and \eqref{steady2} into \eqref{theta1} and \eqref{theta2}, respectively, yields
\begin{align}
\Theta_1=\frac{\psi_1}{\langle k \rangle}\sum_{k'\in\mathcal{K}}\frac{k'^2P(k')\Theta_1}{1+\psi_1k'\Theta_1+\psi_2k'\Theta_2},\label{self1}\\
\Theta_2=\frac{\psi_2}{\langle k \rangle}\sum_{k'\in\mathcal{K}}\frac{k'^2P(k')\Theta_2}{1+\psi_1k'\Theta_1+\psi_2k'\Theta_2}.\label{self2}
\end{align}
Thus, the steady state pair $(\Theta_1^*,\Theta_2^*)$ in (OP2) should satisfy equations \eqref{self1} and \eqref{self2}. 
For clarity, we denote 
\begin{equation}
T_1=\frac{\psi_1\langle k^2 \rangle}{\langle k \rangle},\ T_2=\frac{\psi_2\langle k^2 \rangle}{\langle k \rangle}.
\end{equation}
In general, the ESR for different strains of epidemics are unequal, i.e., $\psi_1\neq \psi_2$. In the special case of 
$\psi_1=\psi_2$, the characteristics of two strains are the same, and it can be seen as a generalized single-strain scenario. Therefore, in the following study, we analyze the network equilibrium in the nontrivial regime $\psi_1\neq\psi_2$.

\subsection{Equilibrium Analysis}\label{equi_analysis}

For the self-consistency equations \eqref{self1} and \eqref{self2}, $(\Theta_1,\Theta_2)=(0,0)$ is a trivial solution. In this case, $\bar{I}_1^*=\bar{I}_2^*=0$ which is a disease-free equilibrium. To obtain nontrivial solutions to \eqref{self1} and \eqref{self2}, we first present the following theorem.

\begin{theorem}\label{thmnopositive}
There exist no positive solutions to the equations \eqref{self1} and \eqref{self2}, i.e., $\Theta_1>0$ and $\Theta_2>0$.
\end{theorem}
\begin{proof}
We proof by contradiction. If there exist positive solutions, i.e., $\Theta_1>0$ and $\Theta_2>0$, \eqref{self1} and \eqref{self2} are equivalent to
\begin{align}
1=\frac{\psi_1}{\langle k \rangle}\sum_{k'\in\mathcal{K}}\frac{k'^2P(k')}{1+\psi_1k'\Theta_1+\psi_2k'\Theta_2},\label{selfreduction1}\\
1=\frac{\psi_2}{\langle k \rangle}\sum_{k'\in\mathcal{K}}\frac{k'^2P(k')}{1+\psi_1k'\Theta_1+\psi_2k'\Theta_2}.\label{selfreduction2}
\end{align}
Since $\psi_1\neq\psi_2$ and $\frac{1}{\langle k \rangle}\sum_{k'}\frac{k'^2P(k')}{1+\psi_1k'\Theta_1+\psi_2k'\Theta_2}>0$, \eqref{selfreduction1} and \eqref{selfreduction2} cannot be satisfied simultaneously which rules out the positive solutions to equations \eqref{self1} and \eqref{self2}. 
\end{proof}

\textit{Remark}: Based on Theorem \ref{thmnopositive}, $\Theta_1$ and $\Theta_2$ cannot be both positive at the steady state, resulting in a \textit{non-coexistence phenomenon} of the two competing strains.

The following corollary on the possible nontrivial solutions of $\Theta_1$ and $\Theta_2$ naturally follows from Theorem \ref{thmnopositive}.
\begin{corollary}\label{possible}
The possible nontrivial solutions to \eqref{self1} and \eqref{self2} fall into two categories: (i) $\Theta_1>0,\Theta_2=0$ and (ii) $\Theta_2>0,\Theta_1=0$. 
\end{corollary}
\begin{proof}
Since $0\leq\Theta_i\leq 1,\ i=1,2$, no negative solutions exist. Then, the possible nontrivial solutions are $\Theta_1>0,\Theta_2=0$ and $\Theta_2>0,\Theta_1=0$. 
\end{proof}

Corollary \ref{possible} indicates that, for the possible nontrivial solutions, strain 1 or strain 2 has an exclusive equilibrium.
The existence of nontrivial solutions is critical for the analysis of network equilibrium. Therefore, we next investigate the conditions under which the network stabilizes at the exclusive equilibrium.

\begin{theorem}\label{exclusivethm}
Strain $i$ has an exclusive equilibrium if and only if $T_i>1$, $i\in\{1,2\}$.
\end{theorem}

\begin{proof}
For the two exclusive equilibria, i.e., $\Theta_1>0,\Theta_2=0$ and $\Theta_2>0,\Theta_1=0$, \eqref{self1} and \eqref{self2} are reduced to 
\begin{equation}\label{reduced_self}
1=\frac{\psi_i}{\langle k \rangle}\sum_{k'}\frac{k'^2P(k')}{1+\psi_i k'\Theta_i},\ i=1,2.
\end{equation}
For the former case $\Theta_1>0,\Theta_2=0$, we define function $g:[0,1]\rightarrow \mathbb{R_+}$, i.e.,
$g(\Theta_1)=\frac{\psi_1}{\langle k \rangle}\sum_{k'}\frac{k'^2P(k')}{1+\psi_1k'\Theta_1}$.
Then, we obtain
\begin{align*}
g(1)=\frac{\psi_1}{\langle k \rangle}\sum_{k'}\frac{k'^2P(k')}{1+\psi_1k'}&=\frac{1}{\langle k \rangle}\sum_{k'}\frac{\psi_1 k'}{1+\psi_1k'} k' P(k')\\
&\hspace{-2ex}< \frac{1}{\langle k \rangle}\sum_{k'}{k'P(k')}=\frac{\langle k \rangle}{\langle k \rangle}=1.
\end{align*}
Moreover,
$
g'(\Theta_1)=-\frac{\psi_1^2}{\langle k \rangle}\sum_{k'}\frac{k'^3P(k')}{(1+\psi_1k'\Theta_1)^2}<0.
$
Therefore, $g$ is a decreasing function over the domain $\Theta_1\in[0,1]$. To ensure the existence of nontrivial solutions to equation \eqref{reduced_self}, a necessary and sufficient condition is $g(0)>1$. Since
$g(0)=\frac{\psi_1}{\langle k \rangle}\sum_{k'}{k'^2P(k')}=\frac{\psi_1\langle k^2 \rangle}{\langle k \rangle}=T_1,$
$g(0)>1$ is equivalent to $T_1>1$.
The analysis is similar for the case $\Theta_2>0,\Theta_1=0$, and the necessary and sufficient condition is $T_2>1$. 
\end{proof}

Three possible equilibria are summarized as follows:
 \begin{itemize}
 \item[1)] Disease-free equilibrium, $E_1=(1,0,0)$.
 \item[2)] Exclusive equilibrium of strain 1, $E_2=(\bar{S}_1^*,\bar{I}_1^*,0)$, if and only if $T_1>1$, where $\bar{S}_1^*$ is the density of healthy nodes at this equilibrium.
 \item[3)] Exclusive equilibrium of strain 2, $E_3=(\bar{S}_2^*,0,\bar{I}_2^*)$, if and only if $T_2>1$, where $\bar{S}_2^*$ is the density of healthy nodes at this equilibrium.
 \end{itemize}
 
\textit{Remark}: $T_i>1$ is equivalent to $\psi_i>\frac{\langle k \rangle}{\langle k^2 \rangle}$, $i=1,2$. In addition, strain $i$ dies out when $\psi_i$ does not satisfy the condition, and the steady state of the network is the disease-free equilibrium $E_1$.

\subsection{Stability Analysis of Equilibria}
In this section, we analyze the stability of the candidate equilibria presented in Section \ref{equi_analysis}.

\begin{theorem}\label{freestable}
If $T_1< 1$ and $T_2< 1$, then the disease-free equilibrium $E_1$ is globally asymptotically stable.
\end{theorem}
\begin{proof}
Note that $\frac{dI_{i,k}(t)}{dt}\leq -\gamma_i I_{i,k}(t)
+\zeta_ik\Theta_i(t)-u_iI_{i,k}(t)$, $i\in\{1,2\}$. Then, it suffices to show that positive solutions of the following auxiliary system $\frac{dI_{i,k}(t)}{dt}= -\gamma_1 I_{i,k}(t)+\zeta_ik\Theta_i(t)-u_iI_{i,k}(t)$ go to zero when $t$ goes to infinity. We define a Lyapunov function $V_i(t):=\sum_{k\in\mathcal{K}}b_k I_{i,k}$, where $b_k = \frac{kP(k)}{\langle k \rangle (u_i+\gamma_i)}$, $i\in\{1,2\}$. Then, we obtain 
\begin{align*}
    \frac{dV_i}{dt} &= \sum_{k\in\mathcal{K}} b_k\left[\zeta_ik\Theta_i(t)-(u_i+\gamma_i)I_{i,k}(t)\right] \\
    & = \sum_{k\in\mathcal{K}} \frac{kP(k)}{\langle k \rangle}\left[ \zeta_ik\Theta_i(t)-(u_i+\gamma_i)I_{i,k}(t)\right]\\
    & = \Theta_i(t)\left[ \frac{\langle k^2 \rangle \zeta_i}{\langle k \rangle(u_i+\gamma_i)}-1\right].
\end{align*}
Thus, when $\frac{\langle k^2 \rangle \zeta_i}{\langle k \rangle(u_i+\gamma_i)}<1$ (which is equivalent to $T_i<1$), then $\frac{dV_i}{dt}<0$, given $\Theta_i\neq 0$, for $i\in\{1,2\}$. We can further conclude that, under $T_i<1$, $\frac{dV_i}{dt}<0$ holds if strain $i$ exists; and $\frac{dV_i}{dt}=0$ only if $I_{i,k}=0$, yielding $\lim_{t\rightarrow\infty}I_{i,k}=0$, $\forall k\in\mathcal{K}$, which is a disease-free equilibrium.
\end{proof}

We further investigate the stability of exclusive equilibrium of strain 1, and the result is presented  as follows.

\begin{theorem}\label{strain1stable}
If $T_1>1$ and $T_1>T_2$, then the exclusive equilibrium of strain 1, $E_2$, is globally asymptotically stable. 
\end{theorem}

\begin{proof}
First, the two coupled non-linear differential equations in \eqref{strains} can be rescaled as  
\begin{equation}\label{strains_normalized}
\begin{split}
\frac{dI_{1,k}(t)}{dt}=& -I_{1,k}(t) +\psi_1k[1-I_{1,k}(t)-I_{2,k}(t)]\Theta_1(t),\\
\frac{dI_{2,k}(t)}{dt}=&- I_{2,k}(t) +\psi_2k[1-I_{1,k}(t)-I_{2,k}(t)]\Theta_2(t).
\end{split}
\end{equation}
The derivative of $\Theta_1(t)$ with respect to time $t$ can thus be given as
$$
\begin{aligned}
\frac{d \Theta_1(t)}{dt} &= \frac{1}{\langle k \rangle}\sum_{k\in \mathcal{K}} k P(k)\frac{dI_{1,k}(t)}{dt}\\
&= \frac{1}{\langle k \rangle}\sum_{k\in \mathcal{K}} k P(k) \left[-I_{1,k} + \psi_1k S_k(t) \Theta_1(t)\right]\\
&=\Theta_1(t)\left[ \frac{1}{\langle k \rangle}\sum_{k\in \mathcal{K}} k P(k)\psi_1 k S_k(t) - 1\right],
\end{aligned}
$$ where $S_k(t):= 1-I_{1,k}(t)-I_{2,k}(t)$.
Similarly, 
$$
\begin{aligned}
\frac{d \Theta_2(t)}{dt}
=\Theta_2(t)\left[ \frac{1}{\langle k \rangle}\sum_{k\in \mathcal{K}} k P(k)\psi_2 k S_k(t) - 1\right].
\end{aligned}
$$
For the exclusive equilibrium of strain $1$, $I_{2,k}^* = 0$ for every $k$ and hence $\Theta_2^* =0$. The steady states at the exclusive equilibrium of strain $1$ agrees with the following identities:
$$
\begin{aligned}
I_{1,k}^* &= \psi_1 k S_k^* \Theta_1^*,\\
S_k^* & =1- I_{1,k}^*,\\
1 &= \frac{1}{\langle k \rangle}\sum_{k\in \mathcal{K}} k P(k)\psi_1 k S_k^*.
\end{aligned}
$$
Consider the following Lyapunov function $V(t)$ for $t\geq 0$, which is defined along a given solution of system \eqref{strains_normalized},
$$
\begin{aligned}
V(t) =& \frac{1}{2}\sum_{k\in\mathcal{K}} \left[ \tilde{b}_k (S_k(t) - S_k^*)^2\right] + \Theta_1(t) - \Theta_1^* \\
&- \Theta_1^* \ln \frac{\Theta_1(t)}{\Theta^*_1}+\Theta_2(t),
\end{aligned}
$$
where the coefficients $\tilde{b}_k >0$ are given by $\tilde{b}_k = \frac{kP(k)}{\langle k \rangle S^*_k}$.
Then the time derivative of $V$ obtained along the solution of system \eqref{strains_normalized} for $t>0$ is
$$
\begin{aligned}
&\frac{dV(t)}{dt} \\
=& \sum_{k\in\mathcal{K}}\left[ \tilde{b}_k (S_k - S^*_k)\frac{dS_k}{dt}\right]+ \frac{\Theta_1 - \Theta_1^*}{\Theta_1}\frac{d\Theta_1}{dt} + \frac{d\Theta_2}{dt}\\
=& \sum_{k\in\mathcal{K}} \tilde{b}_k (S_k-S^*_k) \left[ I_{1,k} + I_{2,k} -\psi_2 k S_k\Theta_2 -\psi_1 k S_k\Theta_1\right]\\
&+(\Theta_1 - \Theta_1^*)\left[ \frac{1}{\langle k \rangle}\sum_{k\in \mathcal{K}} k P(k)\psi_1 k S_k - 1 \right]\\
&+ \Theta_2 \left[ \frac{1}{\langle k \rangle}\sum_{k\in \mathcal{K}} k P(k)\psi_2 k S_k - 1\right]\\
=& \sum_{k\in\mathcal{K}} \tilde{b}_k (S_k - S_k^*)\big[(I_{1,k}-I_{1,k}^*) + I_{2,k}\\ &-\psi_2k(S_k\Theta_2  - S_k^* \Theta_2^*) - \psi_1 k(S_k\Theta_1  - S_k^* \Theta_1^*) \big]\\
&+(\Theta_1-\Theta_1^*) \left[ \frac{1}{\langle k \rangle}\sum_{k\in \mathcal{K}} k P(k)\psi_1 k (S_k-S_k^*) \right]\\
&+ \Theta_2 \left[ \frac{1}{\langle k \rangle}\sum_{k\in \mathcal{K}} k P(k) k (\psi_2 S_k- \psi_1 S_k^*) \right]\\
=& \sum_{k\in\mathcal{K}} \tilde{b}_k \big[(S_k -S_k^*)(I_{1,k} -I_{1,k}^*) +(S_k -S_k^*)I_{2,k}\\
&-\psi_2k \Theta_2(S_k -S_k^*)^2 +\psi_2k S_k^* (S_k-S_k^*)(\Theta_2^* -\Theta_2)\\
&- \psi_1 k\Theta_1(S_k -S_k^*)^2 + \psi_1 k S_k^* (S_k - S_k^*)(\Theta_1^* -\Theta_1) \big]\\
&+\frac{1}{\langle k \rangle}\sum_{k\in \mathcal{K}} k P(k)\psi_1 k (S_k-S_k^*)(\Theta_1-\Theta_1^*)\\
&+\frac{1}{\langle k \rangle}\sum_{k\in \mathcal{K}} k P(k) k (\psi_2 S_k- \psi_1 S_k^*)\Theta_2\\
=&\sum_{k\in\mathcal{K}} \tilde{b}_k \big[ (S_k -S_k^*)(I_{1,k} - I_{1,k}^*) + (S_k - S_k^*)I_{2,k}\\
&-\psi_2 k\Theta_2(S_k-S_k^*)^2 - \psi_1 k \Theta_1(S_k -S_k^*)^2 \big]\\
&- (\psi_1 -\psi_2) \Theta_2\left[ \frac{1}{\langle k\rangle}\sum_{k\in\mathcal{K}}k^2 P(k
)s_k^* \right] \\
\leq &\sum_{k\in\mathcal{K}}\tilde{b}_k \left[(S_k - S_k^*)( 1- S_k -I_{1,k} + I_{1,k} - 1 +S_k^*) \right]\\
&- (\psi_1 -\psi_2) \Theta_2\left[ \frac{1}{\langle k\rangle}\sum_{k\in\mathcal{K}}k^2 P(k
)S_k^* \right]
\leq 0,
\end{aligned}
$$
where the last inequality holds if $\psi_1 \geq \psi_2$, i.e., $T_1 \geq T_2$.
Note that $\frac{dV(t)}{dt} =0$ holds if $S_k = S_k^*$ for $k\in\mathcal{K}$ and 
$\Theta_2^*=0$. Leveraging results from Theorem \ref{exclusivethm}, by LaSalle's invariant principle \cite{khalil2002nonlinear} we can conclude that the exclusive equilibrium of strain $1$ is globally asymptotically stable if $T_1>1$ and $T_1>T_2$. 
\end{proof}

Similarly, we can obtain the condition for stable exclusive equilibrium $E_3$ as follows.

\begin{theorem}\label{strain2stable}
If $T_2>1$ and $T_2>T_1$, then the exclusive equilibrium of strain 2, $E_3$, is globally asymptotically stable. 
\end{theorem}

\begin{proof}
The proof is similar to that in Theorem \ref{strain1stable} and hence omitted here. 
\end{proof}

In Theorems \ref{freestable}, \ref{strain1stable}, and \ref{strain2stable}, the ESR plays an critical role in determining the equilibrium. For example, if ESR of both strains of epidemics are smaller than $\frac{\langle k \rangle}{\langle k^2 \rangle}$, then both epidemics die out at steady state. This disease-free stable state occurs when either the control effort is sufficiently large or the epidemics have a relatively low spreading ability. In comparison, when strain 1's ESR exceeds $\frac{\langle k \rangle}{\langle k^2 \rangle}$ and it is also greater than strain 2's ESR, then only strain 1 exists at equilibrium as shown in Theorem \ref{strain1stable}. This non-coexistence phenomenon indicates that the strain that has a larger spreading rate and is more loosely controlled can eventually survive in the network.

\textit{Remark:} Theorems \ref{freestable}, \ref{strain1stable}, and \ref{strain2stable} provide global convergence guarantees of the non-linear epidemic dynamics, which rules out the possibility of limit cycles as could be observed in other non-linear systems. 

\section{Optimal curing Strategy Design}\label{optimal_control}
We have obtained the stable equilibria of the competing epidemics in Section \ref{equi} which further characterize the steady state expressions of parameters in (OP2). In this section, we aim to determine the optimal curing strategy of epidemics spreading via solving (OP2) in Section \ref{prolem_formulation}.

\subsection{Bounds on Control Effort}\label{control_bound}
Before addressing (OP2), we present the control bounds at each network equilibrium which should be taken into account when designing the optimal control. The following Theorem \ref{Corollary2} directly follows from Theorems \ref{freestable}, \ref{strain1stable}, and \ref{strain2stable}.

\begin{theorem}\label{Corollary2}
The control efforts leading to different network equilibria are summarized as follows.
\begin{enumerate}
\item If the network reaches the disease-free equilibrium $E_1$, the control law needs to satisfy
\begin{align}
u_1> \frac{\zeta_1 \langle k^2 \rangle}{\langle k \rangle}-\gamma_1,\label{boundfree1}\\
u_2> \frac{\zeta_2 \langle k^2 \rangle}{\langle k \rangle}-\gamma_2.\label{boundfree2}
\end{align}
Note that $u_i\geq 0,i=1,2$,  and thus when $\frac{\zeta_i \langle k^2 \rangle}{\langle k \rangle}-\gamma_i < 0, i=1,2$, \eqref{boundfree1} and \eqref{boundfree2} hold.
\item If the network is stabilized at the exclusive equilibrium $E_2$, the control law needs to satisfy
\begin{align}
&u_1<\frac{\zeta_1 \langle k^2 \rangle}{\langle k \rangle}-\gamma_1,\label{boundstrain11}\\
&u_2> \frac{\zeta_2(\gamma_1+u_1)}{\zeta_1}-\gamma_2.\label{boundstrain12}
\end{align}
\item If the network is stabilized at the exclusive equilibrium $E_3$, the control law needs to satisfy
\begin{align}
&u_2<\frac{\zeta_2 \langle k^2 \rangle}{\langle k \rangle}-\gamma_2,\label{boundstrain21}\\
&u_1> \frac{\zeta_1(\gamma_2+u_2)}{\zeta_2}-\gamma_1.\label{boundstrain22}
\end{align}
\end{enumerate}
\end{theorem}

As the results in Theorems \ref{freestable}, \ref{strain1stable}, and \ref{strain2stable} are concerned with the global asymptotic stability, the control bounds presented in Theorem \ref{Corollary2} are sufficient to drive the equilibrium to the desired one. Furthermore, these control bounds in Theorem \ref{Corollary2} have natural interpretations. The efforts to control strains 1 and 2 by the network operator need to be higher than the thresholds shown in \eqref{boundfree1} and \eqref{boundfree2} to achieve a disease-free steady state. In comparison, if only one strain of epidemics exists at the equilibrium, then the control effort to the other strain is upper bounded by a constant as shown in \eqref{boundstrain11} and \eqref{boundstrain21}.

\subsection{Optimal curing of competing Epidemics}\label{optimal_control_epidemics}
In this section, we address the optimal control problem for each equilibrium case.
\subsubsection{Stable disease-free equilibrium}
In this case, the optimization problem (OP2) is reduced to
\begin{align*}
(\mathrm{OP3}):\quad &\min_{\mathbf{u}}\quad c_1(\mathbf{u})+c_2(0)\\
&\mathrm{s.t.}\ \mathrm{inequalities}\ \eqref{boundfree1}\ \mathrm{and}\ \eqref{boundfree2}.
\end{align*}
Note that the solution to (OP3) ensures that the epidemic network will stabilize at the disease-free equilibrium, as reflected in the constraints. The control cost $c_1(\mathbf{u})$ guides the optimal strategy design that can drive the network to such an equilibrium. In the implementation, when two strains of epidemics die out, then the system operator can cease to apply the control.

Due to the monotonicity of function $c_1$, we can obtain the optimal control solutions based on Theorem \ref{Corollary2} as
\begin{equation}\label{free_solution}
\begin{split}
u_1 = \max\left( 0,\frac{\zeta_1 \langle k^2 \rangle}{\langle k \rangle}-\gamma_1\right),\\
u_2= \max\left(0, \frac{\zeta_2 \langle k^2 \rangle}{\langle k \rangle}-\gamma_2\right).
\end{split}
\end{equation}

When $\frac{\zeta_1 \langle k^2 \rangle}{\langle k \rangle}<\gamma_1$ and $\frac{\zeta_2 \langle k^2 \rangle}{\langle k \rangle}<\gamma_2$, then no control is required and the network reaches the disease-free equilibrium automatically at the steady state due to sufficiently high recovery rates $\gamma_1$ and $\gamma_2$ of the epidemics comparing with their spreading rates $\zeta_1$ and $\zeta_2$. We summarize the results of optimal curing at disease-free regime in the following theorem.

\begin{theorem}\label{coro_disease_free}
At the stable disease-free equilibrium, when $\frac{\zeta_1 \langle k^2 \rangle}{\langle k \rangle}<\gamma_1$ and $\frac{\zeta_2 \langle k^2 \rangle}{\langle k \rangle}<\gamma_2$, the optimal  effort is irrelevant with network structure, i.e., the degree distribution $P(k)$, and admits a value 0. When $\frac{\zeta_1 \langle k^2 \rangle}{\langle k \rangle}>\gamma_1$ or $\frac{\zeta_2 \langle k^2 \rangle}{\langle k \rangle}>\gamma_2$, the optimal effort is positive and depends on the average network connectivity $\langle k \rangle$ and the second moment $\langle k^2 \rangle$.
\end{theorem}

\textit{Remark:} In the disease-free regime, Theorem \ref{coro_disease_free} indicates that the optimal curing strategies for networks with different degree distributions $P(k)$ but the same $\langle k \rangle$ and $\langle k^2 \rangle$ are identical, yielding a \textit{distribution independent}  optimal control strategy.

\subsubsection{Stable Exclusive Equilibrium of Strain 1}
Since $\bar{I}_{2,k}^*=0$ in this case, the optimization problem (OP2) becomes
\begin{equation*}
\begin{split}
(\mathrm{OP4}):\quad \min_{\mathbf{u}}\quad &c_1(\mathbf{u})+c_2\left(w_1\bar{I}_1^*(u_1)\right)\\
\mathrm{s.t.}\quad &{I}_{1,k}^*(u_1)=\frac{\psi_1k\Theta_1^*}{1+\psi_1k\Theta_1^*},\ \forall k\in \mathcal{K},\\
&\psi_1=\zeta_1/(\gamma_1+u_1),\\
&\mathrm{inequalities}\ \eqref{boundstrain11}\ \mathrm{and}\ \eqref{boundstrain12},
\end{split}
\end{equation*}
where $\Theta_1^*$ and $\bar{I}_1^*(u_1)$ are defined in (OP2). Similar to (OP3), (OP4) also belongs to a subcase towards solving (OP2). We will comment on a specific mechanism in determining the optimal solution to (OP2) in the end of this section.

To solve (OP4), we obtain an expression of $\bar{I}_1^*(u_1)$ with respect to $u_1$. Note that $\bar{I}_{1,k}^*(u_1), k\in\mathcal{K}$, and $\Theta_1^*$ are coupled in the constraints, and we need to solve the following system of equations:
\begin{align}
&{I}_{1,k}^*(u_1)=\frac{\psi_1k\Theta_1^*}{1+\psi_1k\Theta_1^*},\ k\in\mathcal{K},\label{coupled1}\\
&\Theta_1^*=\frac{\sum_{k'}k'P(k')I_{1,k'}^*(u_1)}{
\langle k \rangle}.\label{coupled2}
\end{align}

To address this problem, we substitute \eqref{coupled1} into \eqref{coupled2} and arrive at the following fixed-point equation:
\begin{align}
\Theta_1^*=\frac{1}{\langle k \rangle}\sum_{k'}\frac{k'^2P(k')\psi_1\Theta_1^*}{1+\psi_1k'\Theta_1^*}.\label{fixeqn}
\end{align}

For the existence and uniqueness of the solutions to \eqref{fixeqn}, we have the following proposition.

\begin{proposition}\label{uniquesolution}
There  exists a unique non-trivial solution $\Theta_1^*$ to the fixed-point equation \eqref{fixeqn}.
\end{proposition}

\begin{proof}
From the proof of Theorem \ref{exclusivethm}, we know that function $g(\Theta_1)=\frac{\psi_1}{\langle k \rangle}\sum_{k'}\frac{k'^2P(k')}{1+\psi_1k'\Theta_1}$ is monotonously decreasing over the domain $\Theta_1\in[0,1]$. Moreover, $g(0)>1$ if $\psi_1 > \langle k \rangle/\langle k^2 \rangle $ and $g(1)<1$ for all possible $\psi_1$. Therefore, $g(\Theta_1)=1$ has a non-trivial solution over $\Theta_1\in (0,1)$, and the solution is unique. 
\end{proof}

\textit{Remark:} The existence and uniqueness of $\Theta_1^*$ ensures the \textit{predictability} of ${I}_{1,k}^*(u_1)$ through \eqref{coupled1}.

Another critical aspect of (OP4) is the continuity of $\bar{I}_1^*(u_1)$ with respect to $u_1$. When $\bar{I}_1^*(u_1)$ is continuous with $u_1$, the objective function in (OP4) is a continuous convex function, and thus can be theoretically solved by using the first-order optimality condition directly. When $\bar{I}_1^*(u_1)$ encounters jumps at some points of $u_1$,  which is a possible case, (OP4) becomes challenging to solve, since $c_2(w_1\bar{I}_1^*(u_1))$ is discontinuous in $u_1$. If this possible discontinuity feature is neglected, the obtained optimal control law is incorrect. To rule out the probability of discontinuity case, we have the following result.

\begin{proposition}\label{continuousmap}
In the optimization problem (OP4), the mapping $\bar{I}_1^*(u_1)$ is continuous in $u_1$.
\end{proposition}

\begin{proof}
The proof to show that $\bar{I}_1^*(u_1)$ is continuous follows similar arguments as in the proof of Berge's maximum theorem. Define $H(x,y):=\frac{1}{\langle k \rangle}\sum_{k'}\frac{k'^2P(k')x y}{1+k'x y}$, where $x\in \mathcal{X}:= (\frac{\langle k \rangle}{\langle k^2 \rangle },\infty)$ and $y\in \mathcal{Y}:=(0,1)$. Note that the fixed-point solution of $y = H(x,y)$ is a solution of \eqref{fixeqn} given $x=\psi_1$. It is easy to see that $H$ is continuous over $\mathcal{X} \times \mathcal{Y}$. Define $f(x,y) = \Vert H(x,y) - y \Vert$, where $\Vert \cdot\Vert$ is a proper norm. Since norms are continuous, $f(x,y)$ is continuous over $\mathcal{X}\times \mathcal{Y}$. From Proposition \ref{uniquesolution}, we know that for any given $x\in\mathcal{X}$, there exists a unique minimizer $y_x\in \mathcal{Y}$ that minimizes $f(x,y)$. Then there exists a map $x \mapsto y_x$ denoted by $h:\mathcal{X}\rightarrow \mathcal{Y}$.  To show $\bar{I}_1^* (u_1)$ is continuous in $u_1$, it is sufficient to show that the map $h$ is continuous. Now we show the map $x\mapsto y_x$ is indeed continuous.  Suppose that the map is not continuous. Then we can find an $\epsilon >0$ such that for all $\delta >0$, there is an $x\in\mathcal{X}$, such that $\Vert x - x'\Vert <\delta $ but $\Vert h(x) - h(x') \Vert >\epsilon $. That means there is an $x'$ such that $\Vert x-x' \Vert <\delta$ and some $\epsilon'>0$ such that $\Vert f(x,h(x)) - f(x',h(x)) \Vert = \Vert f(x,h(x)) + f(x',h(x')) - f(x',h(x))\Vert > \epsilon'$, where we use the fact that $f(x,h(x)) = f(x',h(x'))=0$ and $\min_{y\in\mathcal{Y}}f(x',y)$ admits a unique minimizer. This contradicts the fact that $f(\cdot,y)$ is continuous over $\mathcal{X}$. Hence, the map $h$ is continuous.  With a slight abuse of notion, the map $\Theta_1(\psi_1)$ that defines the solution of \eqref{fixeqn} is continuous in $\psi_1$. From the definition of $\bar{I}_1^*$ in \eqref{I_1_def} and the definition of $\psi_1$ in \eqref{psi_def}, we can conclude that $\bar{I}_1^*(u_1)$ is continuous in $u_1$, which completes the proof.
\end{proof}

\textit{Remark:} Based on Proposition \ref{continuousmap}, the continuous mapping $\bar{I}_1^*(u_1)$ leads to a robust epidemic control scheme. Specifically, with a small perturbation of the unit control cost (e.g., a small change on the constants $K_1$ and $K_2$ when the control cost function admits a form of $c_1(\mathbf{u})=K_1u_1+K_2u_2$), the severity of epidemics under the optimal control resulting from (OP4) does not encounter a significant deviation. The reason is that a perturbation on the unit control cost (e.g., $K_1$ or $K_2$ above) does not yield a large change on the optimal control and thus $\bar{I}_1^*(u_1)$ based on its continuity.

To obtain the solution $\Theta_1^*$ with respect to $\psi_1$, we first denote the right hand side of \eqref{fixeqn} as a function of $\Theta_1^*$, i.e, $Q:\ [0,1]\rightarrow [0,1]$. Specifically, 
\begin{equation}
Q(\Theta_1^*)=\frac{1}{\langle k \rangle}\sum_{k'}\frac{k'^2P(k')\psi_1 \Theta_1^*}{1+\psi_1k'\Theta_1^*}.
\end{equation}

Then, \eqref{fixeqn} can be solved by using the following fixed-point iterative scheme:
\begin{align}
\Theta_1^{*(n+1)}=Q(\Theta_1^{*(n)}),\ n=0,1,2,...,N,
\end{align}
until $\vert Q(\Theta_1^{*(n+1)})-\Theta_1^{*(n+1)} \vert \leq \epsilon_1$, where $\epsilon_1>0$ is the predefined error tolerance. The algorithm to obtain solution $\Theta_1^{*}$ is summarized in Algorithm \ref{algorithm1}.

The convergence of the fixed-point iterative scheme is guaranteed which is summarized in the following result.

\begin{lemma}
The iterative scheme in Algorithm \ref{algorithm1} converges to the unique fixed-point solution.
\end{lemma}

\begin{proof}
Let $H(x,y)$ be the function defined in the proof of Proposition \ref{continuousmap} over the same domain $\mathcal{X}\times \mathcal{Y}$. Define a fixed-point iteration as $y_{k+1} = H(x,y_{k})$. We show that the sequence $\{y_k,k\in\mathbb{N}\}$ converges to the fixed-point solution $y_x$ with $y_x = H(x,y_x)$ for any given initial point $y_0\in \mathcal{Y}$. First, we have
\begin{align*}
\frac{\partial}{\partial y}H(x,y)&=\frac{1}{\langle k \rangle}\sum_{k'}\frac{k'^2x P(k')(1+k'x y)-k'^3P(k')x^2 y}{(1+k'x y)^2}\\
&=\frac{1}{\langle k \rangle}\sum_{k'}\frac{k'^2 x P(k') }{(1+k'x y)^2}>0,
\end{align*}
That means for any given $x$, $H(x,y)$ is monotonically increasing over $y$. Then for any given $y_0\in \mathcal{Y}$, $y_0\neq y_x$, we have $y_1>y_0$ or $y_1 <y_0$. If $y_1 > y_0$, then $H(x,y_1) > H(x,y_0)$, which means $y_2 >y_1$. Then the sequence $\{y_k,k\in\mathbb{N}\}$ is monotonically increasing. Similarly, we can show that if $y_1<y_0$, the sequence  $\{y_k,k\in\mathbb{N}\}$ is monotonically decreasing. Also, it is easy to see that $H(x,y)\in (0,1)$ for all $(x,y)\in\mathcal{X},\mathcal{Y}$. Hence, the sequence $\{y_k,k\in\mathbb{N}\}$ is bounded. From proposition \ref{uniquesolution}, we know for any given $x\in\mathcal{X}$, the  fixed-point solution is achieved within $(0,1)$ and unique. By monotone convergence theorem, we can conclude that the sequence $\{y_k,k\in\mathcal{K}\}$ converges to $y_x$.
\end{proof}

For a given $\Theta_1^{*}$, we have $\bar{I}_{1}(u_1)=\sum_k P(k) I_{1,k}(u_1)$, where ${I}_{1,k}(u_1)=\frac{\zeta_1k\Theta_1^{*}}{\gamma_1+u_1+\zeta_1k\Theta_1^{*}}$. Define a function $f:\mathbb{R}_+^2\rightarrow\mathbb{R}_+$ by
\begin{equation}
f(\mathbf{u}):=c_1(\mathbf{u})+c_2\big(w_1\bar{I}_1(u_1)\big).
\end{equation}
Since $\mathbf{u}\geq 0$, $c_2\big(w_1\bar{I}_1(u_1)\big)$ is continuously differentiable, and so does $f(\mathbf{u})$.  To minimize $f(\mathbf{u})$, we use the gradient descent method incorporating with backtracking line search to obtain the optimal control $\mathbf{u}^*$.

For clarity, the complete proposed method is summarized in Algorithm \ref{algorithm3}.

\begin{algorithm}[t]
\caption{Fixed-Point Iterative Scheme}\label{algorithm1}
\begin{algorithmic}[1]
\State Initialize $\Theta_1^{*(0)},\ \epsilon_1,\ n=0$
\State Calculate $Q(\Theta_1^{*(n)})$
\While {$\vert Q(\Theta_1^{*(n)})-\Theta_1^{*(n)} \vert >\epsilon_1$}
\State$\Theta_1^{*(n+1)}=Q(\Theta_1^{*(n)})$
\State $n=n+1$
\EndWhile
\State \textbf{return} $\Theta_1^{*(n)}$
\end{algorithmic}
\end{algorithm}

\begin{algorithm}[t]
\caption{Gradient Descent Method based on Fixed-Point Iterative Scheme}\label{algorithm3}
\begin{algorithmic}[1]
\State Initialize the starting point $u^{(0)}=0$, $n=0$, tolerance $\epsilon_2$, $u^{(-1)}=\epsilon_2+1$. Obtain a feasible set $\mathcal{U}$ of effort $\mathbf{u}$ from \eqref{boundstrain11} and \eqref{boundstrain12} 
\While {$||\mathbf{u}^{(n)}-\mathbf{u}^{(n-1)}||_2>\epsilon_2$}
\State $\psi_1^{(n)}=\frac{\zeta_1}{\gamma_1+u_1^{(n)}}$
\State Obtain value $\Theta_1^{*(n)}$ through Algorithm \ref{algorithm1}
\For {$k=0 : K$}
\State ${I}_{1,k}(u_1)=\frac{\zeta_1k\Theta_1^{*(n)}}{\gamma_1+u_1+\zeta_1k\Theta_1^{*(n)}}$
\EndFor
\State $\bar{I}_{1}(u_1)=\sum_k P(k) I_{1,k}(u_1)$.
\State Obtain $\mathbf{u}^*=\mathrm{arg}\min\limits_{\mathbf{u}} c_1(\mathbf{u})+c_2\Big(w_1\bar{I}_1(u_1)\Big)$ using gradient descent method
\State $\mathbf{u}_f^*=\mathrm{Proj}_{\mathcal{U}} (\mathbf{u}^*)$
\State $n=n+1$
\State $\mathbf{u}^{(n)}=\mathbf{u}_f^*$
\EndWhile
\State \textbf{return} $\mathbf{u}_f^*$
\end{algorithmic}
\end{algorithm}

\subsubsection{Stable Exclusive Equilibrium of Strain 2}
Since $\bar{I}_{1,k}^*=0$, the optimization problem (OP2) becomes
\begin{equation*}
\begin{split}
(\mathrm{OP5}):\quad \min_{\mathbf{u}}\quad &c_1(\mathbf{u})+c_2\left(w_2\bar{I}_2^*(u_2)\right)\\
\mathrm{s.t.}\quad &{I}_{2,k}^*(u_2)=\frac{\psi_2k\Theta_2^*}{1+\psi_2k\Theta_2^*},\ \forall k\in \mathcal{K},\\
&\psi_2=\zeta_2/(\gamma_2+u_2),\\
&\mathrm{inequalities}\ \eqref{boundstrain21}\ \mathrm{and}\ \eqref{boundstrain22},
\end{split}
\end{equation*}
where $\Theta_2^*$ and $\bar{I}_2^*(u_2)$ are presented in (OP2). Since (OP5) is similar to (OP4),  the analysis to obtain the optimal control $\mathbf{u}^*$ also follows and is omitted here.

We next comment on one observation of the optimal control effort with respect to the network structure. Different from the distribution independent strategy in disease-free regime where $\langle k \rangle$ and $\langle k^2 \rangle$ are sufficient statistics, the node degree distribution $P(k)$ plays an essential role in the optimal control of epidemics in the exclusive equilibria of strain 1 and strain 2. We summarize this result in the following corollary.

\begin{corollary}
In the exclusive equilibria of strain 1 and strain 2 regime, the optimal control effort is distribution dependent, i.e., correlated with the node degree distribution $P(k)$, $\forall k\in\mathcal{K}$, as the epidemic severity cost $c_2$ depends on the average epidemic level including all nodes' degree classes.
\end{corollary}

\textit{Remark:} We have characterized the best curing strategy in each equilibrium regime. The next critical problem is to characterize the global optimal strategy across three equilibria, which is the solution to the original problem (OP2). This goal can be achieved as follows. After obtaining each optimal curing strategy corresponding to different equilibria, we then compare the objective values in these three cases. The one associated with the lowest cost among $(\mathrm{OP3})$, $(\mathrm{OP4})$, and $(\mathrm{OP5})$ is the global optimal strategy.

Note that if the system operator has a predefined goal of the steady state of the network, then it is sufficient to solve one of the problems $(\mathrm{OP3})$, $(\mathrm{OP4})$, and $(\mathrm{OP5})$. In such scenarios, the designed control is \textit{regime-aware} by taking the control bounds in Section \ref{control_bound} into account.

\section{Equilibria Switching via Optimal curing}\label{switching_section}
In this section, we present a switching phenomenon of network equilibria. Specifically, when the equilibrium state of the epidemic network without control effort is not disease-free, then it can switch to different equilibrium states through the applied control effort. To better illustrate this phenomenon, we focus on a class of symmetric control schemes and the system operator aims to suppress two epidemics jointly.  Furthermore, we consider the nontrivial case $\frac{\zeta_1}{\gamma_1}\neq \frac{\zeta_2}{\gamma_2}$ where two strains of epidemics are distinguishable.

\subsection{Motivation of Equilibria Switching}
Before presenting the formal results, we provide an intuitive example to motivate this switching phenomenon. Recall that the optimal effort depends on the tradeoff between the epidemic severity cost and the control cost captured by $c_1$ and $c_2$, respectively. Then, the steady state of epidemic network  can switch if the unit cost of control effort changes. For example, the control cost of strain 1 is relatively high at the beginning which prohibits the system operator in adopting $u_1$ and thus the anticipated network equilibrium only contains strain 1. However, the control cost of strain 1 may decrease significantly due to the maturity of curing technology for agents infected by strain 1, and thus control effort $u_1$ can be applied to suppress the epidemic spreading before its outbreak. The increase of $u_1$ may lead to an equilibrium switching from $E_2$ to $E_3$ as the total cost of network with steady state $E_3$ is lower than the one stabilized at $E_2$, and hence it is an optimal strategy for the system operator.

\subsection{Symmetric Control Effort Scenario}
 In general, $u_1$ and $u_2$ can admit different values. For ease of presenting the structural results, we  focus on the symmetric control scenario $u_1=u_2=u$ and comment on the general case later in this section. This scenario is practical as the global system operator aims to suppress the spreading of two strains simultaneously. In addition, the unit cost of control effort of two strains decreases, and thus the optimal effort $u$ increases continuously based on the continuity result in Proposition \ref{continuousmap}.
 Depending on the parameters of the epidemics, the increasing optimal control can lead to either \textit{single} or \textit{double} switching between equilibrium points. Based on Theorems \ref{freestable}, \ref{strain1stable}, and \ref{strain2stable}, we obtain the following corollary which presents the conditions under which the network encounters a single switching of equilibria. 

\begin{corollary}\label{single_transition}
Consider the case that $\frac{\zeta_i \langle k^2 \rangle}{\gamma_i \langle k \rangle}>1$, and $\frac{\zeta_i}{\gamma_i}>\frac{\zeta_{-i}}{\gamma_{-i}}$, where $i=1\ \mathrm{or}\ 2$, and $-i:=\{1,2\}\setminus\{i\}$, i.e., the epidemic network is stabilized at the exclusive equilibrium of strain $i$ without control. If
$$\zeta_i\geq \zeta_{-i}\quad \textit{or}$$
$$\zeta_i<\zeta_{-i}\ \ \mathrm{and}\ \ {\zeta_i-\gamma_i}>{\zeta_{-i}-\gamma_{-i}},$$ 
then, there exists a single transition from the exclusive equilibrium of strain $i$ to the disease-free equilibrium with the increase of optimal $u$.
\end{corollary}
The single switching phenomenon in Corollary \ref{single_transition} enhances the \textit{prediction} of network equilibrium under control, since it confirms that the exclusive equilibrium of strain $-i$ is not possible under the symmetric optimal control case in this parameter regime.

Similarly, the phenomenon of double switching of equilibrium points is presented as follows.

\begin{corollary}\label{double_transition}
Consider the case that $\frac{\zeta_i \langle k^2 \rangle}{\gamma_i \langle k \rangle}>1,\ i=1,2$, i.e., the epidemic network does not reach the disease-free equilibrium without control. When 
$$\frac{\zeta_i}{\gamma_i}>\frac{\zeta_{-i}}{\gamma_{-i}},\ \ \zeta_i<\zeta_{-i},\ \ \mathrm{and}\ \ \frac{\zeta_i-\gamma_i}{\zeta_{-i}-\gamma_{-i}}<1,$$
where $i=1,2$ and $-i:=\{1,2\}\setminus \{i\}$, then, there exist transitions from the exclusive equilibrium of strain $i$, to the exclusive equilibrium of strain $-i$, and to the disease-free equilibrium with the increase of $u$.
\end{corollary}

For the special case that $\frac{\zeta_1}{\gamma_1}= \frac{\zeta_2}{\gamma_2}$, and $\frac{\zeta_i \langle k^2 \rangle}{\gamma_i \langle k \rangle}>1,\ i=1,2$, when $\zeta_i>\zeta_{-i}$, there exist transitions from the current network equilibrium (mixed steady state with both strains) to the exclusive equilibrium of strain $i$, and then to the disease-free equilibrium with the increase of optimal control $u$ as the unit control cost decreases.

To identify the optimal policies under which the control effort leads to a stable disease-free equilibrium through switching, we present the following definition.

\begin{definition}[Fulfilling Threshold]\label{fulfilling_definition}
The fulfilling threshold refers to the optimal control $\bar{\mathbf{u}}=(\bar{u}_1,\bar{u}_2)$ under which the epidemic network stabilizes at the disease-free equilibrium after switching of network equilibria, and the total cost $c_1(\bar{\mathbf{u}})$ is the lowest among all control policies. Equivalently, $\bar{\mathbf{u}}$ satisfies the following conditions:
\begin{equation}
\begin{split}
c_1(\bar{\mathbf{u}})\leq c_1(\mathbf{u}),\ \forall \mathbf{u},\\
\bar{u}_1> \frac{\zeta_1 \langle k^2 \rangle}{\langle k \rangle}-\gamma_1,\\
\bar{u}_2> \frac{\zeta_2 \langle k^2 \rangle}{\langle k \rangle}-\gamma_2.
\end{split}
\end{equation}
\end{definition}

Based on Definition \ref{fulfilling_definition}, we next characterize the fulfilling threshold in the investigated scenario.

\begin{proposition}\label{fulfilling_prop}
The optimal control effort does not increase after the epidemic network switches from the exclusive equilibrium $E_2$ or $E_3$ to the disease-free equilibrium $E_1$. In the investigated symmetric control scenario with constraint $u_1=u_2$, the fulfilling threshold is 
\begin{equation}\label{fulfilling_u}
\bar{u} = \max\left(0,\frac{\zeta_1 \langle k^2 \rangle}{\langle k \rangle}-\gamma_1,\frac{\zeta_2 \langle k^2 \rangle}{\langle k \rangle}-\gamma_2\right).
\end{equation}
\end{proposition}
 
\begin{proof}
The fulfilling threshold in the studied cases can be directly verified by the zero epidemic cost in regime $E_1$ and the monotonically increasing function  $c_1$ with respect to the applied effort. Based on Definition \ref{fulfilling_definition} and symmetric control structure, we can obtain the threshold $\bar{u}$ in \eqref{fulfilling_u}.
\end{proof}

\textit{Remark:} The fulfilling threshold in Proposition \ref{fulfilling_prop} provides an upper bound for the network operator's control effort to bring the network equilibria to the disease-free regime. As the unit cost of effort decreases, the amount of optimal control should not exceed the fulfilling threshold.

Another result on the number of network equilibria switching is summarized as follows.

\begin{corollary}\label{coro_num_switch}
Under the symmetric optimal control scenario with decreasing unit control cost, the maximum number of network equilibria switching is two.
\end{corollary}

Corollary \ref{coro_num_switch} generalizes Corollaries \ref{single_transition} and \ref{double_transition} by studying the entire parameter regime. The monotonically increasing optimal control yields either single or double switching of equilibria. For general cases in which optimal $u_1$ and $u_2$ are not necessarily the same, then the switching of network equilibria depends on the specific unit costs of $u_1$ and $u_2$. However, if the system operator has a preference to avoid the outbreak of strain $i$, then as the optimal control $u_i$ increases, either single or double switching happens with the network stabilizing at disease-free equilibrium depending on the epidemic system parameters.

\section{Numerical Experiments}\label{simulation}
In this section, we corroborate the obtained results with numerical experiments. First, we generate a scale-free (SF) network with 500 nodes using the Barab{\'a}si-Albert model \cite{barabasi1999emergence}. The SF model has been found very successful in capturing the features and properties of a large number of real-world networks, including social networks, computer networks, financial networks, and airline networks, etc \cite{barabasi2003linked}. Comparing with random homogeneous network and small world network, the SF network is more suitable for our applications of epidemics spreading in social networks and viruses spreading in the Internet, as these real-world networks usually exhibit the characteristics that a significant number of nodes have a lot of connections (seen as hubs), and a trailing tail of nodes have a few connections. Further, as claimed in \cite{barabasi2003linked}, power laws on the degree distribution (namely SF network) are ubiquitous in complex networks. More detailed comparisons between these models as well as the practical examples of SF network modeling can be found in  \cite{wang2003complex}.

In our numerical experiment, the degree distribution of the SF network satisfies $P(k)\sim k^{-3}$. The typical generated random network in the following studies has an average connectivity $\langle k \rangle =1.996$ and $\langle k^2 \rangle =13.75$.  Our objective is to design the optimal control of competing epidemics spreading under different network equilibrium cases. During control implementation, increasing the curing/recovery rate can be achieved, for instance, by allocating antidotes or providing other forms of treatment to a fraction of the vulnerable population (e.g., the infected population or likely infected population). Each infected node is treated homogeneously with a same probability of receiving treatment in the complex network. Note that the curing rate is increased when additional resources are leveraged to facilitate the recovery process of a fraction of population. The functions in the optimization problems admit the forms: $c_1(\mathbf{u})=K_1u_1+K_2u_2$, and $c_2(w_1\bar{I}_1^*(u)+w_2\bar{I}_2^*(u))=K_3(\bar{I}_1^*(u)+\bar{I}_2^*(u))$, where $K_1$, $K_2$ and $K_3$ are positive constants, and $w_1=w_2=1$. Specifically, we choose $K_1=15$, $K_2=10$ and $K_3=50$. For better illustration purposes, we assume that strain 1 and strain 2 have the same spreading rate, i.e., $\zeta_1=\zeta_2=\zeta$.  We find and compare the optimal control solutions of the following two scenarios: scenario I where $\gamma_1=0.5,\gamma_2=0.3$, and  scenario II where $\gamma_1=0.5,\gamma_2=0.8$.

\subsection{Optimal Control in Disease-Free Case}\label{study_free}
\begin{figure}[t]
  \centering
  \subfigure[optimal control ($u_1$ overlaps in two cases)]{%
    \includegraphics[width=0.5\columnwidth]{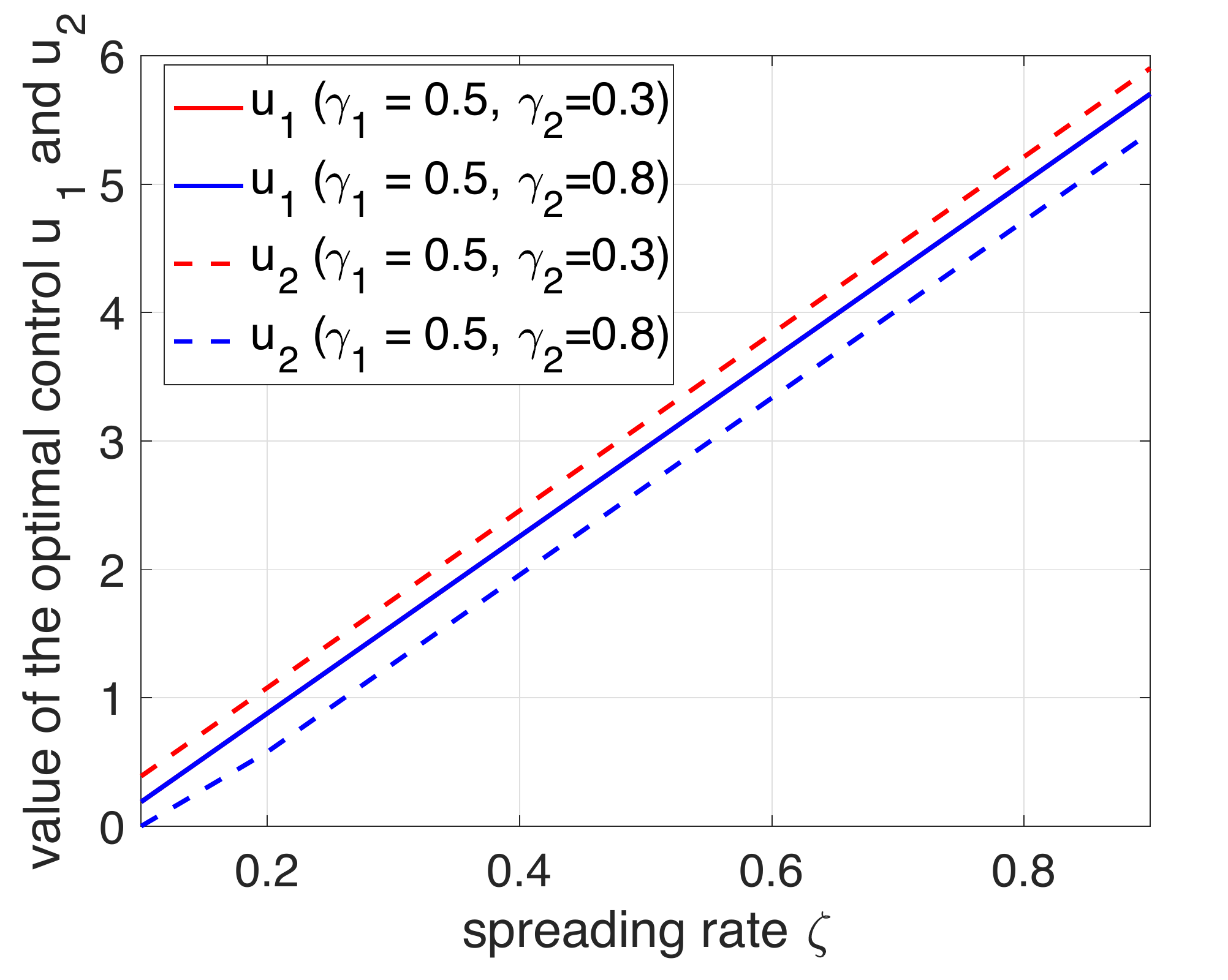}%
    \label{fressresults_1}%
    }%
	 \subfigure[objective value]{%
    \includegraphics[width=0.5\columnwidth]{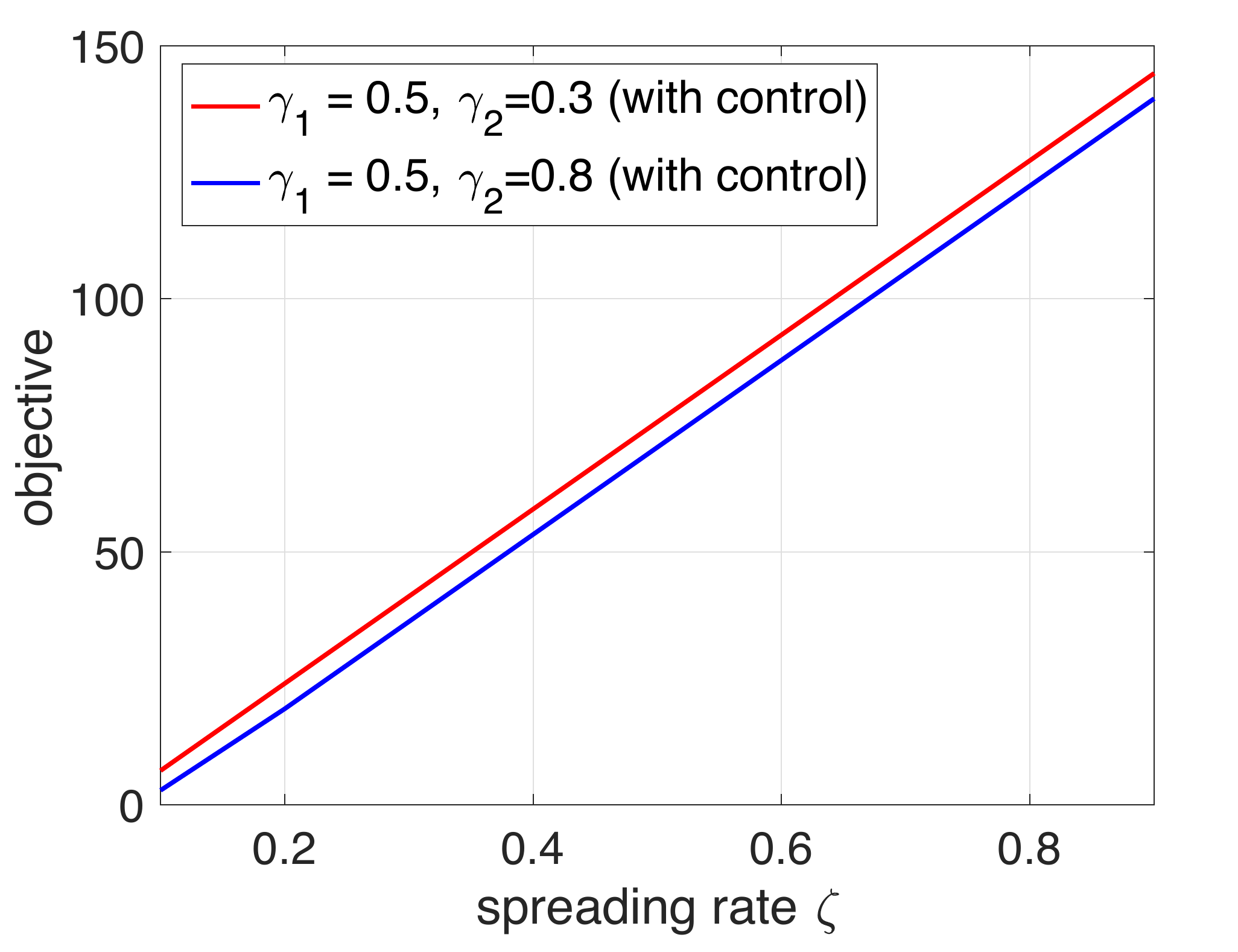}%
    \label{fressresults_2}%
    }
  \caption[]{(a) and (b) show the results of the optimal control and the associated objective value, respectively, where the network stabilizes at the disease-free equilibrium.}
  \label{fressresults}
\end{figure}

In the disease-free case, the epidemic spreading levels are zero at the steady state. By solving (OP3), we obtain the results of optimal control which are shown in Fig. \ref{fressresults}. We can see that the control efforts $u_1$ and $u_2$ both increase linearly with the spreading rate $\zeta$ as expected by \eqref{free_solution}. Due to the same recovery rates of strain 1 in two scenarios, the applied control efforts $u_1$ overlap as shown in Fig. \ref{fressresults_1}. In addition, because of a smaller self-recovery rate of strain 2 in scenario I, its corresponding control effort $u_2$ is larger than that in scenario II. Hence, the optimal objective value in scenario II is smaller than that of scenario I shown in Fig. \ref{fressresults_2}.

\subsection{Optimal Control in Exclusive Equilibrium Case}\label{study_exclusive}
\begin{figure}[t]
  \centering
  \subfigure[]{%
    \includegraphics[width=0.25\textwidth]{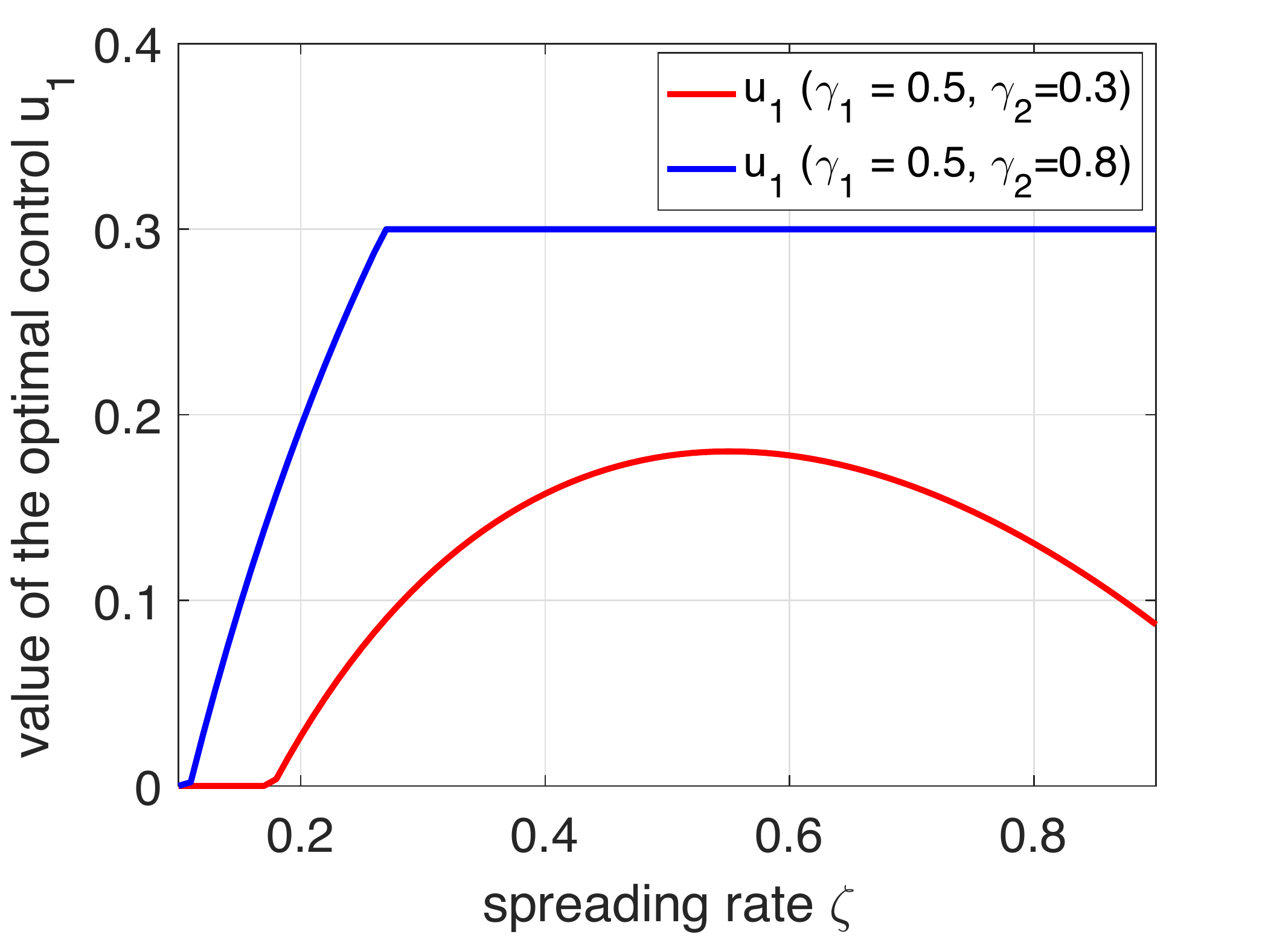}%
    \label{linear_u1}%
  }%
  \subfigure[]{%
    \includegraphics[width=0.25\textwidth]{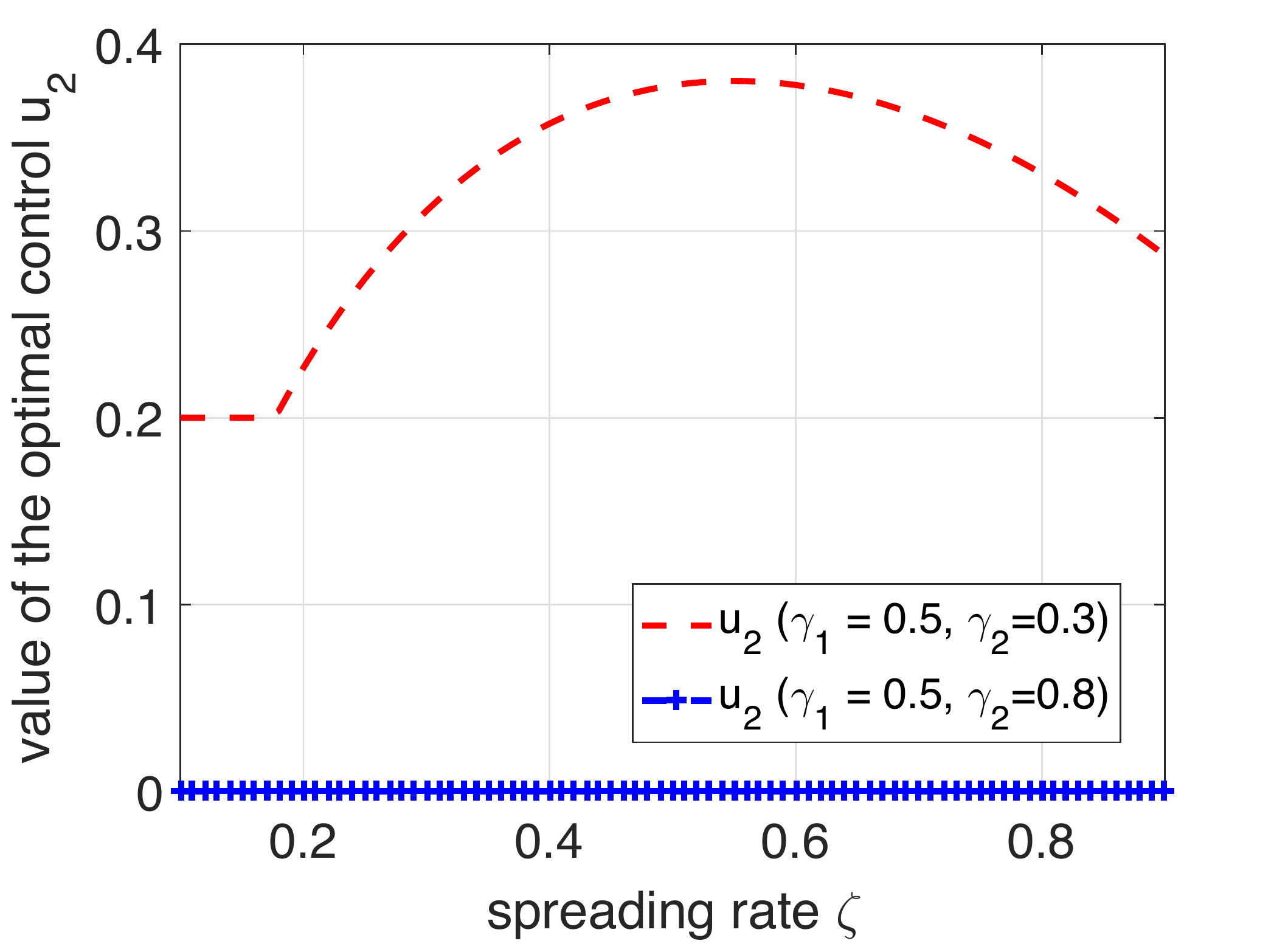}%
    \label{linear_u2}%
  }%
  \hfill
  \subfigure[]{%
    \includegraphics[width=0.25\textwidth]{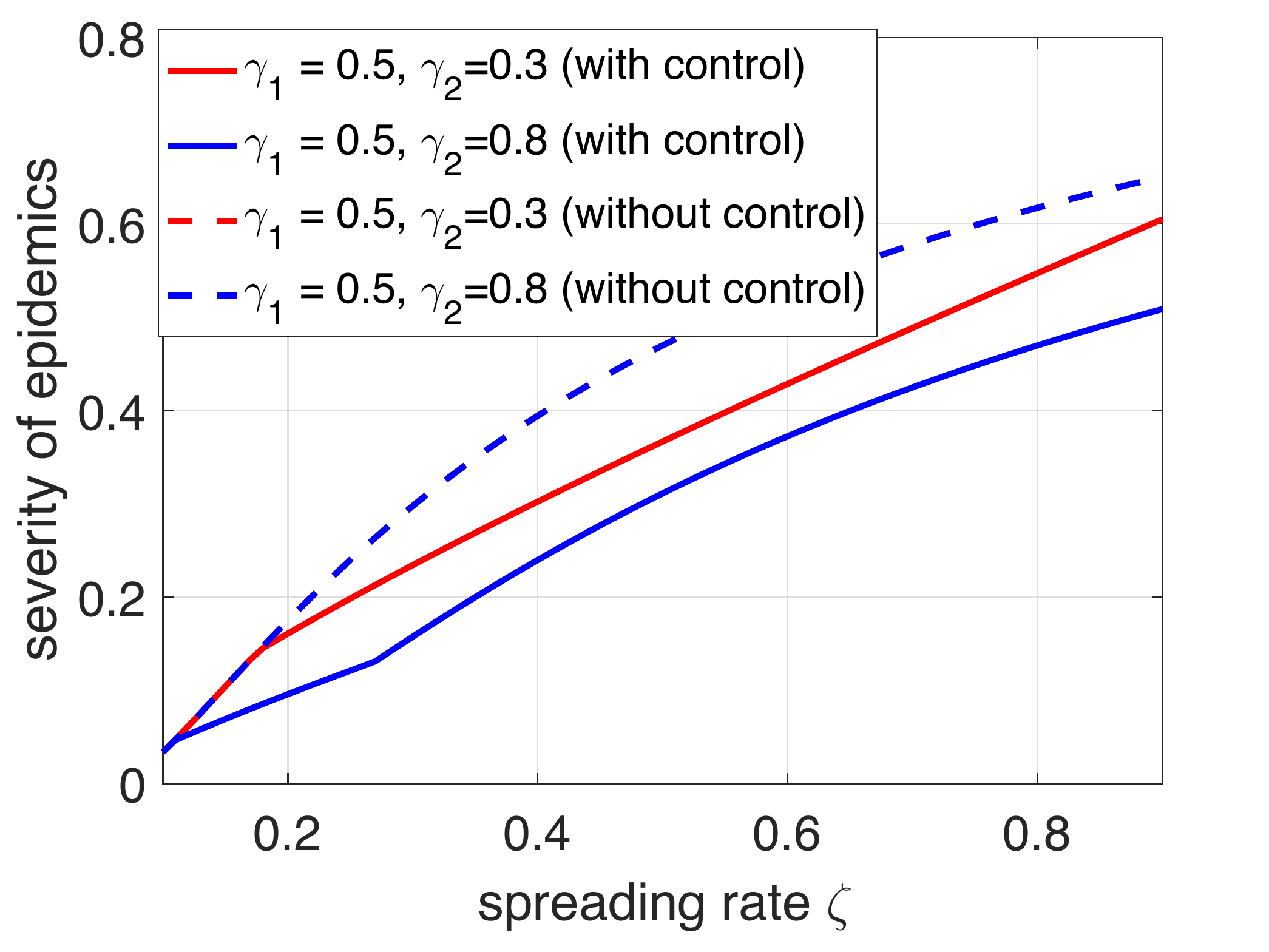}%
    \label{linear_epidemic}%
  }%
  \subfigure[]{%
    \includegraphics[width=0.25\textwidth]{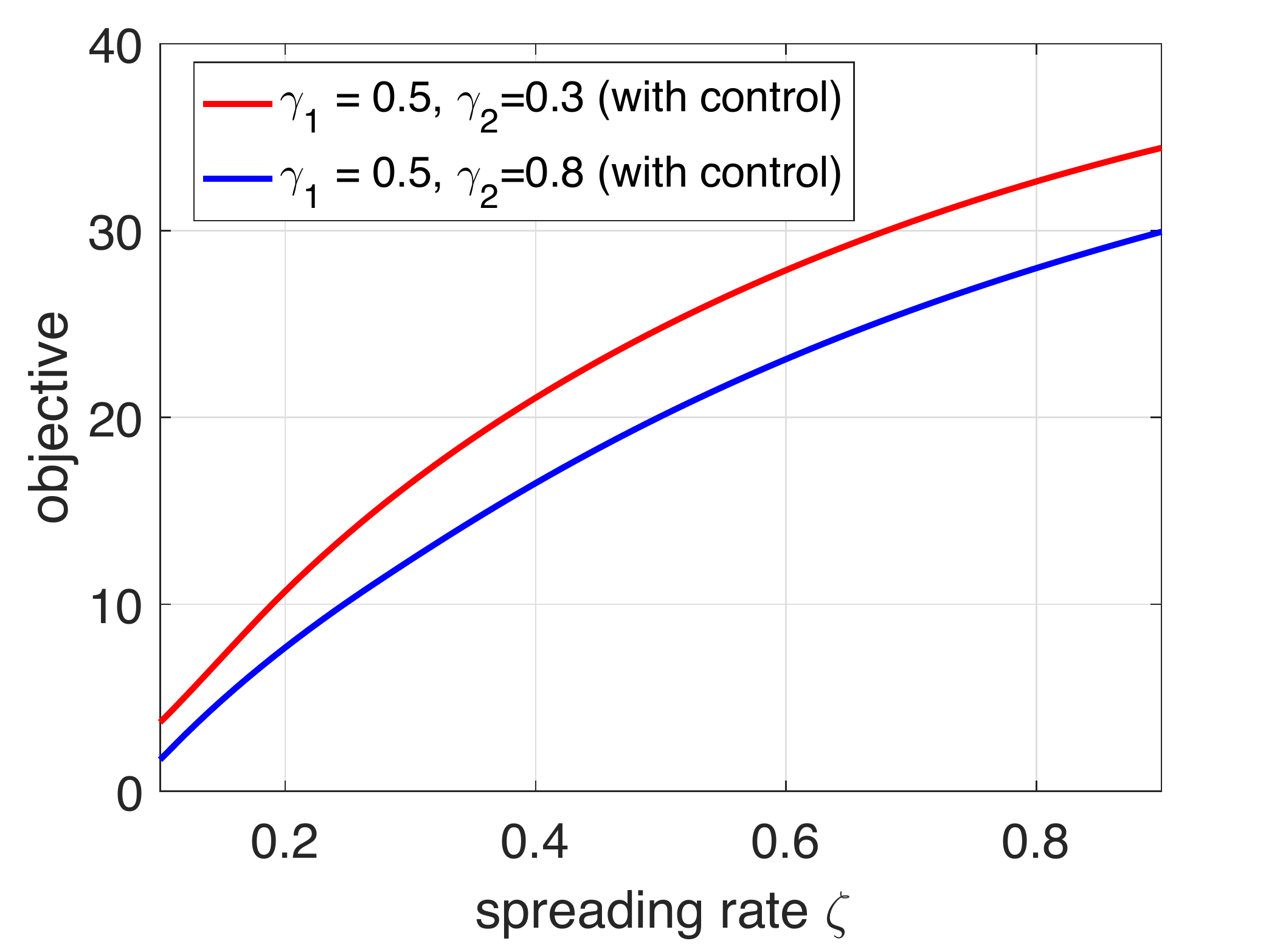}%
    \label{linear_obj}%
  }%
  \caption{The network is stabilized at the exclusive equilibrium of strain 1. (a) and (b) are the optimal control of strain 1 and strain 2, respectively. (c) and (d) show the severity of epidemics and the corresponding objective value under the optimal control, respectively.}
  \label{bigfig1}
\end{figure}

We investigate the case when the network is stabilized at the exclusive equilibrium of strain 1. By solving (OP4) using the proposed Algorithm \ref{algorithm3}, the obtained results are shown in Fig. \ref{bigfig1}. Specifically, Figs. \ref{linear_u1} and \ref{linear_u2} show the optimal control efforts. In scenario I, the control $u_1$ (red line in Fig. \ref{linear_u1}) increases first when the spreading rate $\zeta$ is relatively small. It then decreases after $\zeta>0.55$, since it is not economical to control the spreading of strain 1 comparing with its control cost.  Further, because the recovery rate of strain 2 in scenario I is low, the applied control $u_2$ (red dotted line in Fig. \ref{linear_u2}) should be relatively large to suppress its spreading. An important phenomenon is that $u_2$ decreases after $\zeta>0.55$, which follows the pattern of $u_1$, since $u_2$ can be chosen as long as it satisfies the conditions in Theorem \ref{strain1stable}, and strain 2 does not exist at the steady state. In scenario II, due to the high self-recovery rate $\gamma_2$, strain 2 dies out at the equilibrium even without control. Thus, the control of strain 2 is 0, i.e., $u_2=0$ (blue dotted line in Fig. \ref{linear_u2}). In addition, the control $u_1$ in this scenario (blue line in Fig. \ref{linear_u1}) first increases to compensate the spreading of strain 1. Then, it stays flat after $\zeta>0.27$, since otherwise larger control $u_1$ leads to a network equilibrium switching from $E_2$ to ${E}_3$. Fig. \ref{linear_epidemic} depicts the severity of epidemics at the steady state with and without control. We can conclude that the optimal control effectively reduces the spreading of epidemics in both scenarios. Note that the epidemic spreading levels without the control intervention overlap in two cases (dotted lines in Fig. \ref{linear_epidemic}) though only strain 2 and strain 1 exist at equilibrium in scenarios I and II, respectively. The reason is that the severity of epidemics is determined by the network structure and the steady state, while the parameter $\zeta$ only influences the rate of epidemics spreading.

\subsection{Transition of the Equilibrium through Control}
\begin{figure}[t]
  \centering
    \includegraphics[width=0.8\columnwidth]{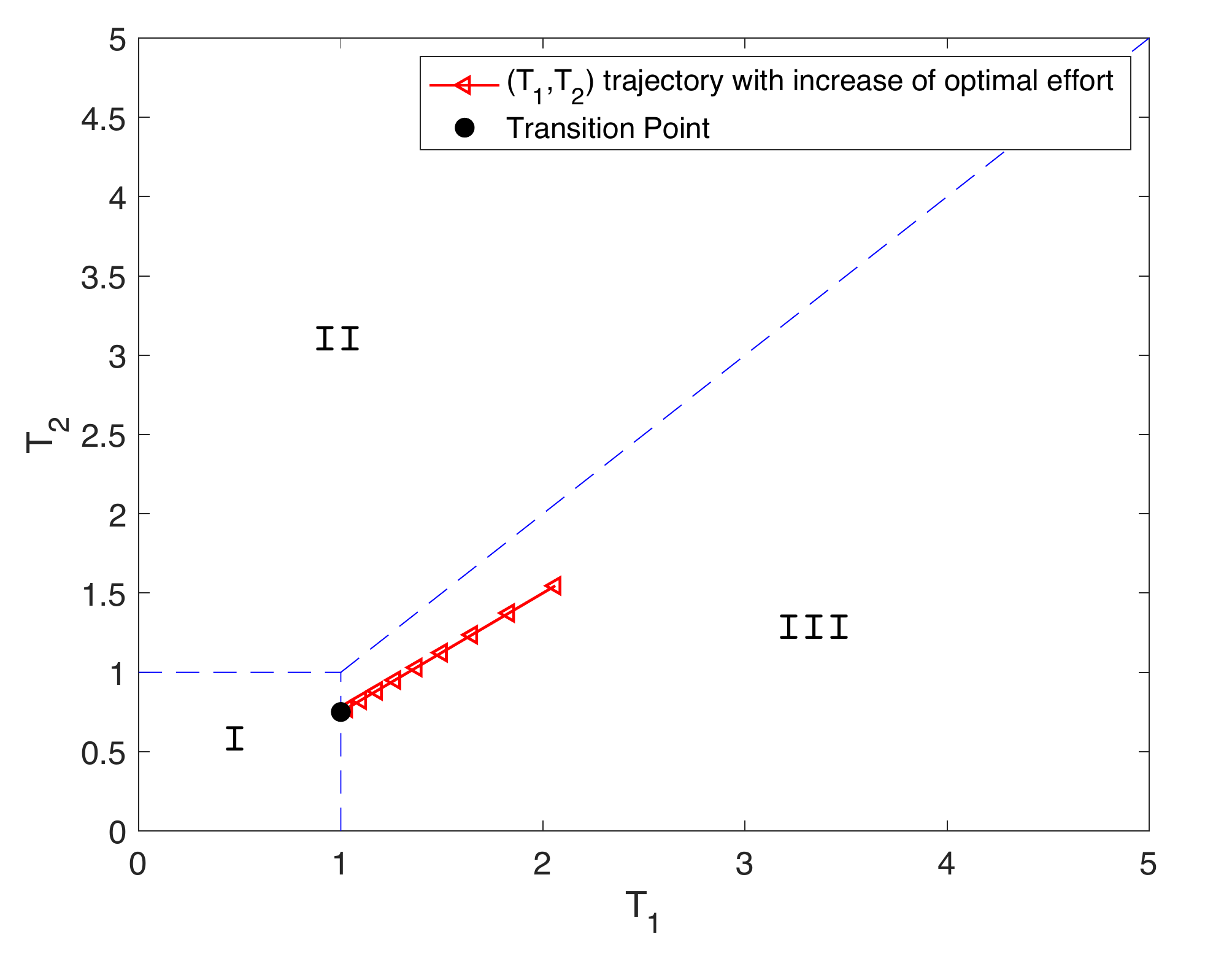}
  \caption[]{Transition of the equilibrium with the increase of control across two regimes: from the exclusive equilibrium of strain 2 (III) to the disease-free equilibrium (I). }\label{transition1}
\end{figure}

\begin{figure}[t]
  \centering
    \includegraphics[width=0.8\columnwidth]{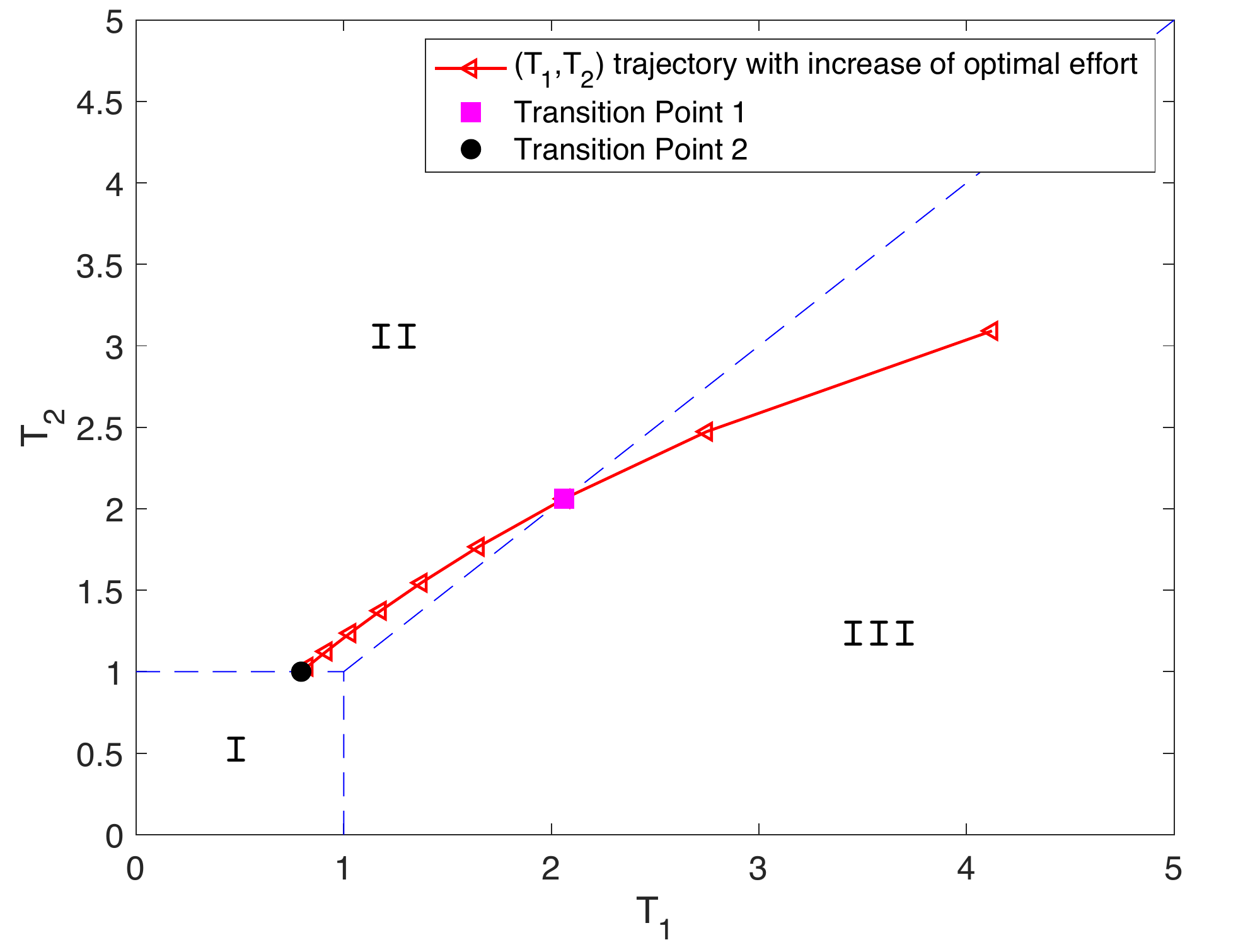}
  \caption[]{Transition of the equilibrium with the increase of control across three regimes. (from the exclusive equilibrium of strain 2 (III) to the exclusive equilibrium of strain 1 (II), then to the disease-free equilibrium (I)).}\label{transition2}
\end{figure}

In this section, we illustrate the transition between the epidemic equilibrium through control.
First, we study the single transition case. From Corollary \ref{single_transition}, we choose $\zeta_1=0.2,\gamma_1=0.4,\zeta_2=0.15$ and $\gamma_2=0.4$. The result is shown in Fig. \ref{transition1}. As the unit control cost changes, the network equilibrium at steady state will be different. Specifically, as the optimal control increases due to the decrease of unit control cost, the epidemic network equilibrium switches from the exclusive equilibrium of strain 2 to the disease-free equilibrium.  For the double transitions case, based on Corollary \ref{double_transition}, we select parameters $\zeta_1=0.1,\gamma_1=0.1,\zeta_2=0.15$ and $\gamma_2=0.2$. The result is shown in Fig. \ref{transition2}. Consistent with Corollary \ref{double_transition}, the network equilibrium switches first from the exclusive equilibrium of strain 2 to the exclusive equilibrium of strain 1, and then to the disease-free equilibrium, as the applied optimal control increases. One common feature in these two cases is that once the effort drives the network to the disease-free equilibrium, the control effort ceases to increase,  where fulfilling threshold is reached (corresponding to the effort level at the transition point denoted by black dot in Figs. \ref{transition1} and \ref{transition2}). Specifically, based on Proposition \ref{fulfilling_prop}, the fulfilling thresholds in Figs. \ref{transition1} and \ref{transition2} are 0.978 and 0.834, respectively.

\begin{figure}[t]
  \centering
  
  \subfigure[]{%
    \includegraphics[width=0.5\columnwidth]{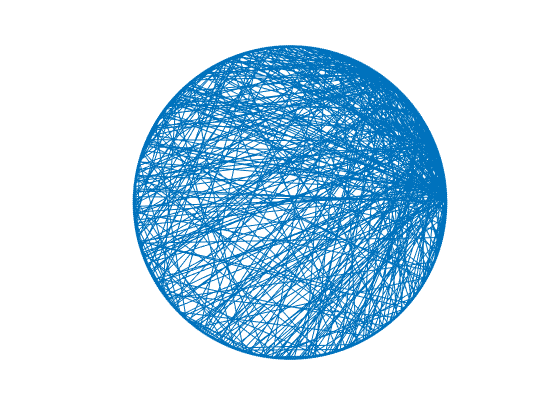}%
    \label{net1}%
  }%
  \hfill
  \subfigure[]{%
    \includegraphics[width=0.5\columnwidth]{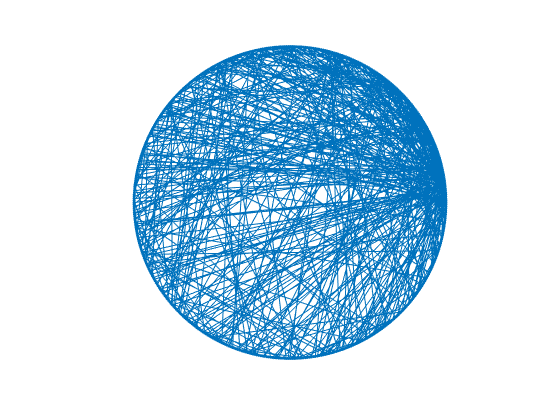}%
    \label{net1_1}%
  }%
  \hfill
  \subfigure[]{%
    \includegraphics[width=0.5\columnwidth]{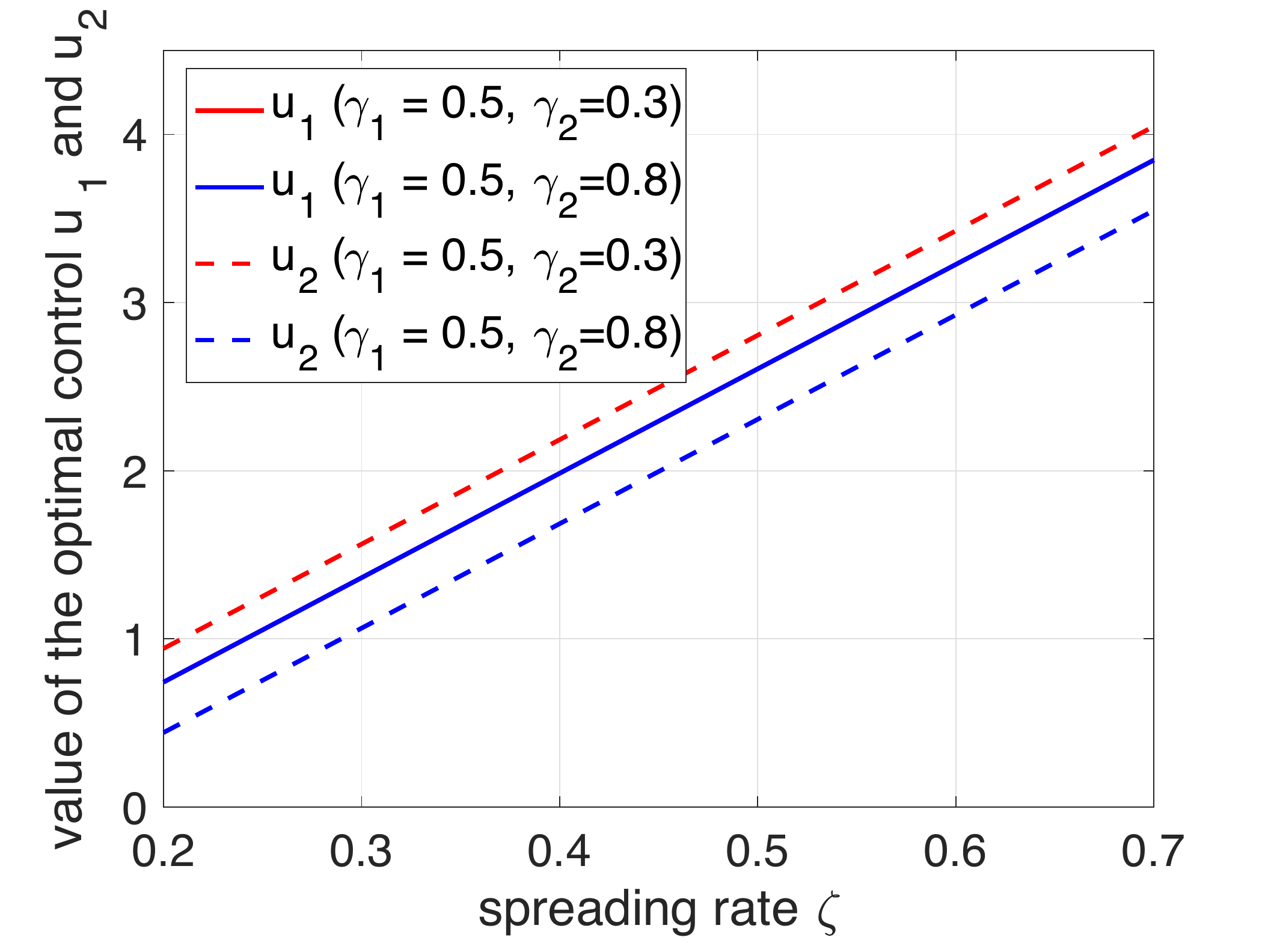}%
    \label{Net2_1}%
  }%
  \hfill
  \subfigure[]{%
    \includegraphics[width=0.5\columnwidth]{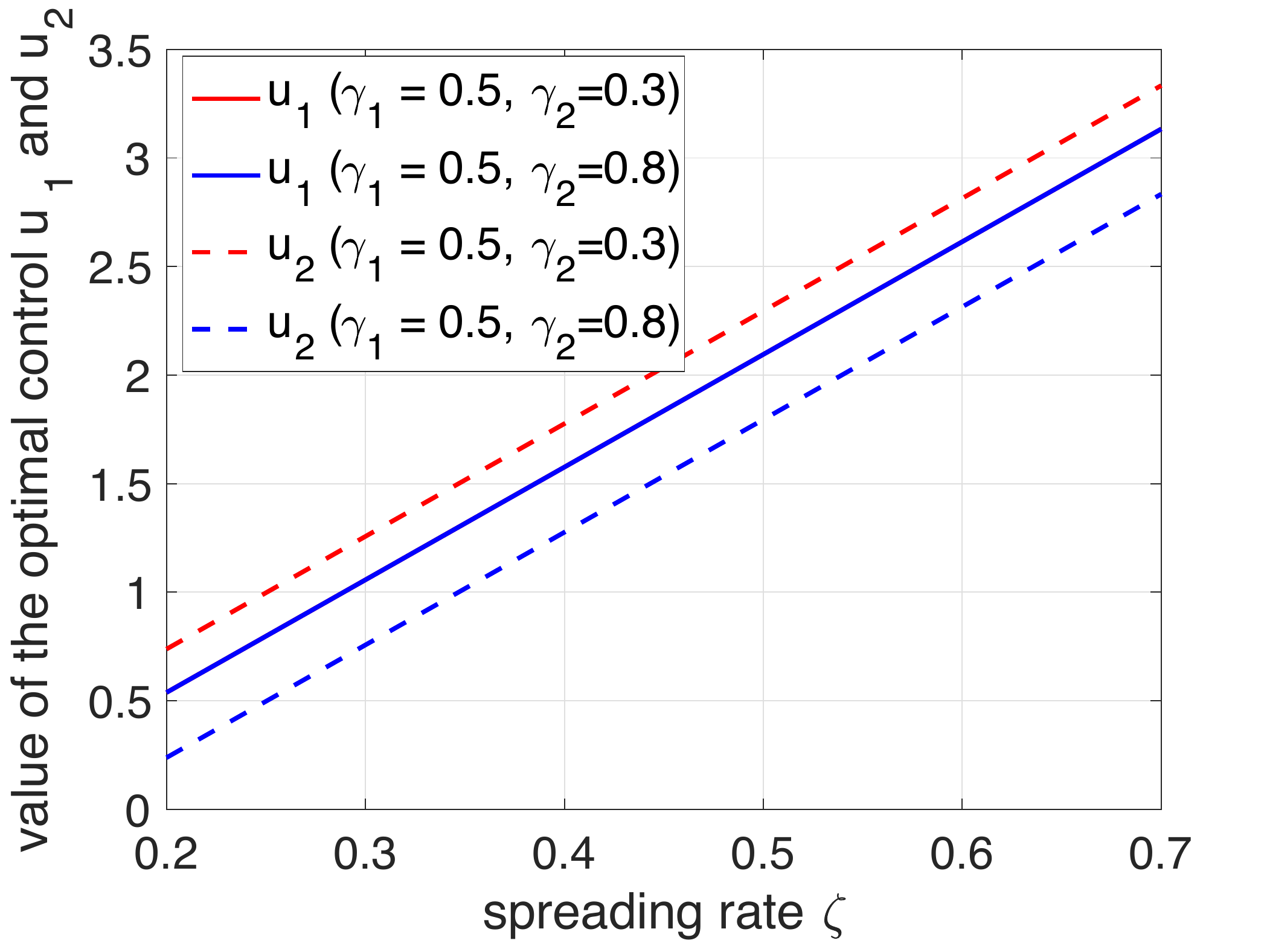}%
    \label{Net2_2}%
  }%
  \hfill
  \subfigure[]{%
    \includegraphics[width=0.5\columnwidth]{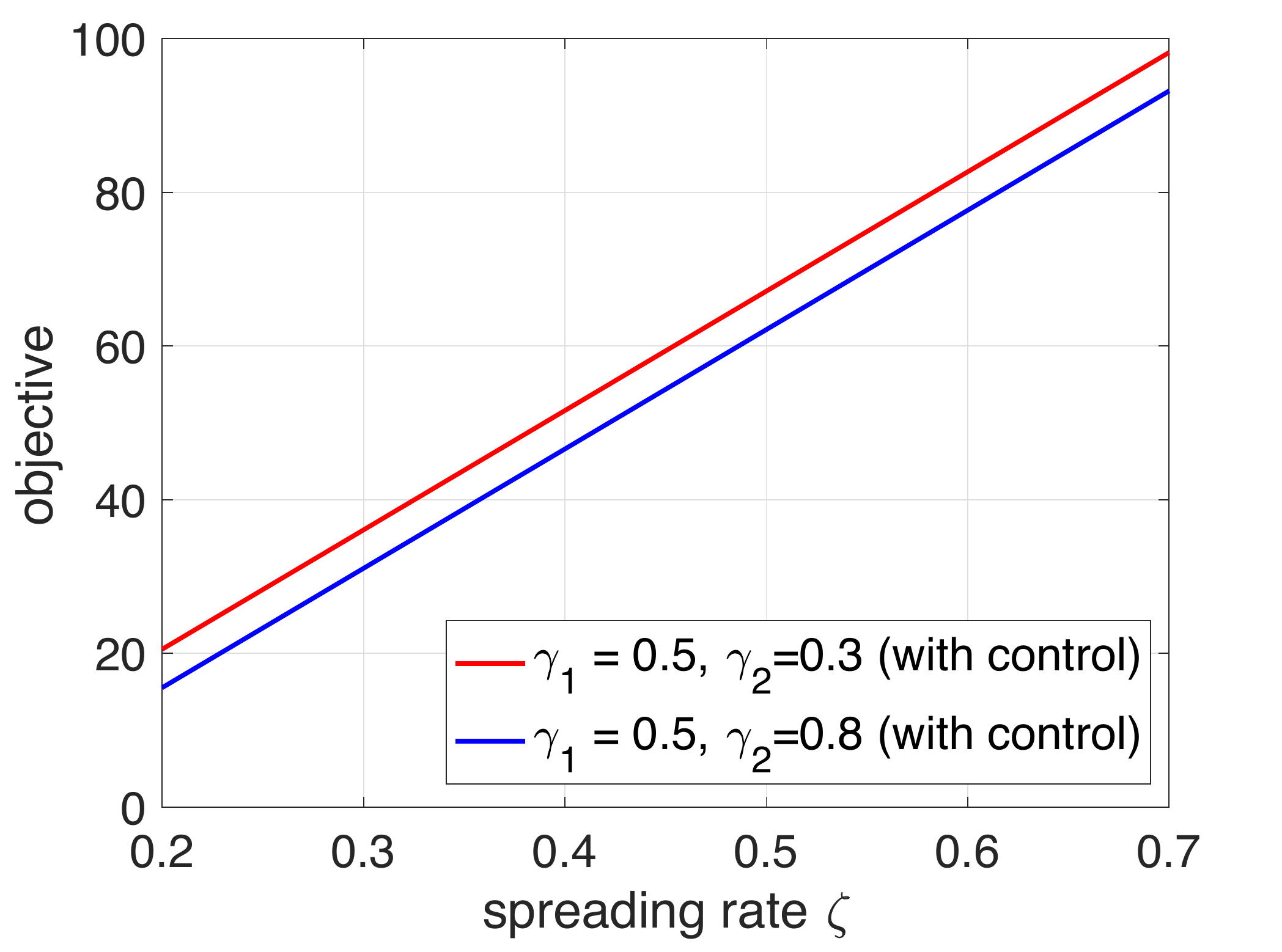}%
    \label{net1_2}%
  }%  
  \hfill
  \subfigure[]{%
    \includegraphics[width=0.5\columnwidth]{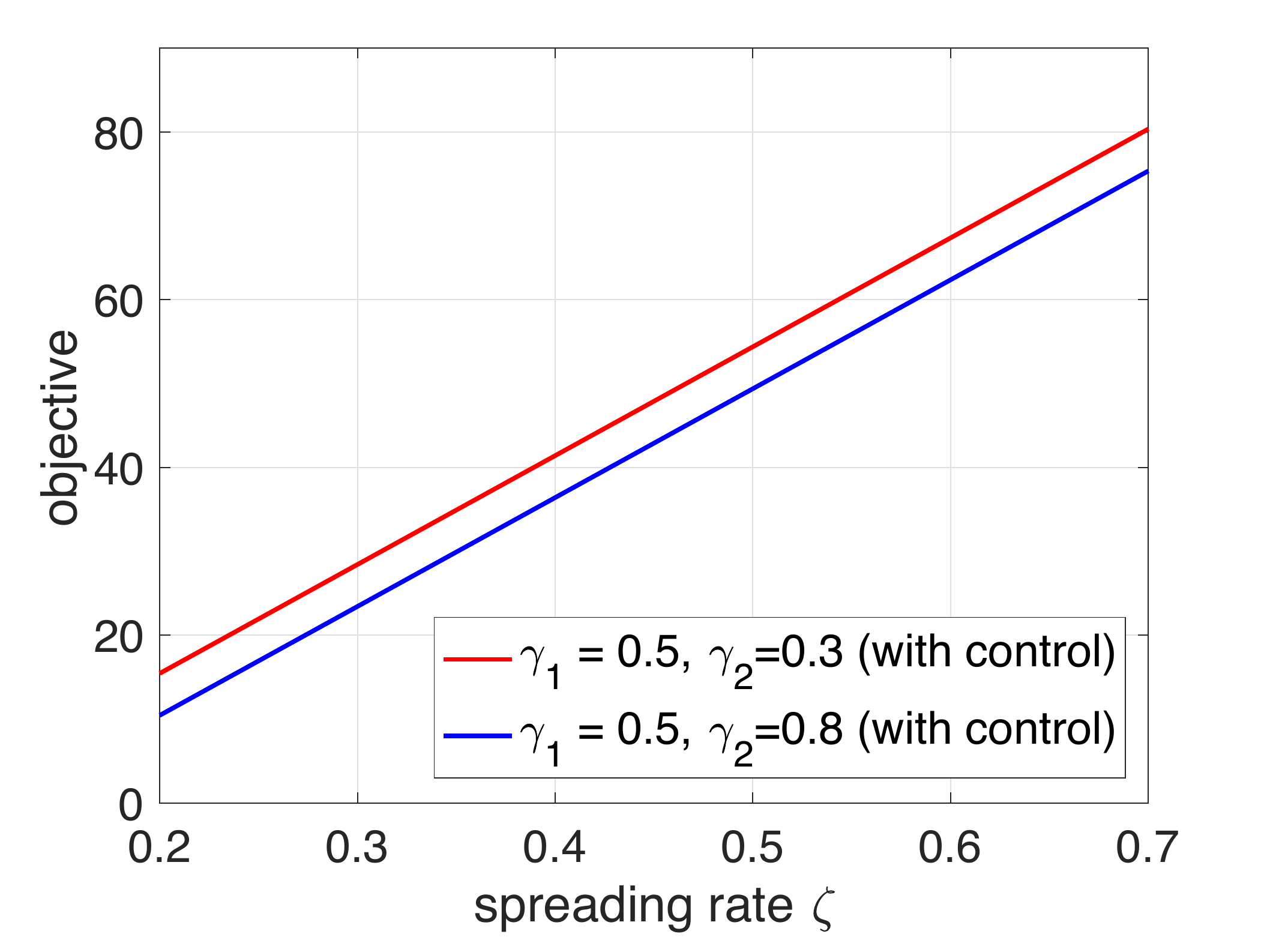}%
    \label{net2}%
  }%
  \caption{(a) and (b) present two random realizations of SF network. (c) and (e) show the results of the network in (a), while (d) and (f) show the results of network in (b). Both networks in (a) and (b) stabilize at the disease-free equilibrium. The differences on the control and the objective in two cases are relatively small.}
  \label{ex:random}
\end{figure}

\subsection{Network Rewiring and Model Mismatch}
\begin{figure}[t]
  \centering
  \subfigure[Optimal control designed for network in Fig. \ref{net1} and applied it to network in Fig. \ref{net1} and Fig. \ref{net1_1} separately.]{%
    \includegraphics[width=0.75\columnwidth]{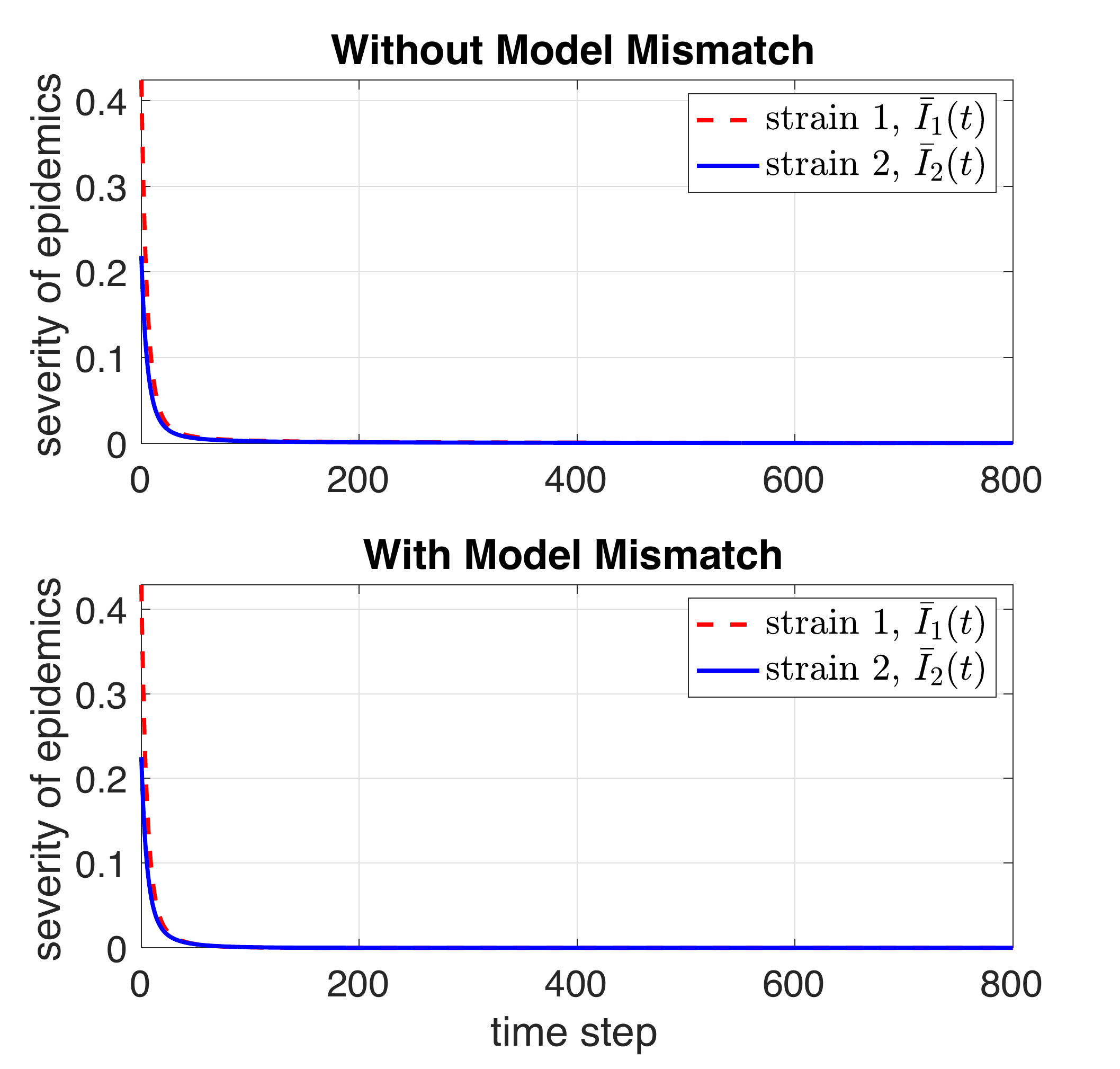}%
    \label{model_mismatch_2}%
  }%
  \hfill
  \subfigure[Optimal  control designed for network in Fig. \ref{net1_1} and applied it to network in Fig. \ref{net1_1} and Fig. \ref{net1} separately.]{%
    \includegraphics[width=0.75\columnwidth]{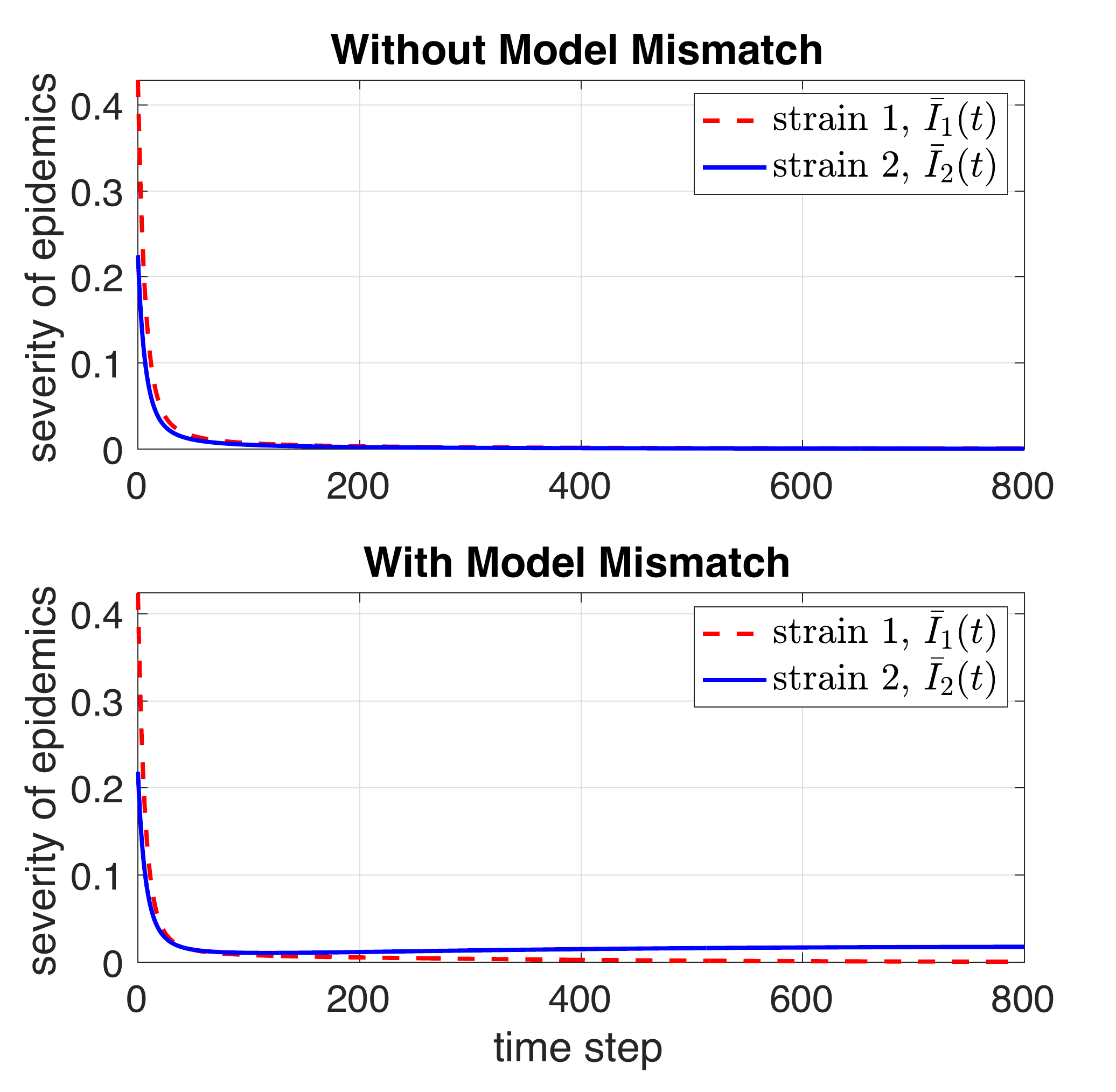}%
    \label{model_mismatch_1}%
  }%
  \caption{(a) and (b) present the results when the system operator has and has not model uncertainty on the epidemic network. In both (a) and (b), the epidemics extinct under the designed optimal control without model mismatch. However, in (b), strain 2 will exist at the equilibrium state if the system operator has an inaccurate network model. In contrast, the network still achieves the disease-free equilibrium in (a) under model mismatch. Thus, a more conservative control design is preferred to eliminate the viruses when the system operator has uncertainty on the underlying epidemic network structure.}
  \label{fig:model_mismatch}
\end{figure}

In this subsection, we compare the results of our model under different random realizations of networks. We generate two SF networks with 500 nodes from the same generative model and revisit the disease-free case. The results of experiments are shown in Fig. \ref{ex:random}. Note that network in (a) has $\langle k^2 \rangle = 12.3960$ while network in (b) has $\langle k^2 \rangle = 10.3600$ and both have a same level of average connectivity $\langle k \rangle = 1.996$. Network (b) could be considered as a network that changes from network (a) with randomly rewired network connections, or the degree-based mean-field model has some mismatch with the real network. We can see that the control efforts in network (b) are slightly lower than the one in network (a), and the optimal objective values in network (b) are also slightly lower than the ones in network (b). We next comment on the consequences when the system operator has inaccurate modeling of the epidemic network. Specifically, consider that the real network admits a structure of network (a), but the system operator assumes that it is of network (b) and designs the corresponding control. Then, based on the optimal solution presented in \eqref{free_solution}, the designed control is not sufficient to achieve a disease-free equilibrium, and the network stabilizes at one of the exclusive equilibria (the exact equilibrium depends on $\gamma_1$ and $\gamma_2$). In comparison, if the real network admits a structure of network (b) but the system operator assumes that it is of network (a), then the designed corresponding control is still able to drive the system to a disease-free equilibrium, showing the robustness of the control in such scenarios. The reason is that the designed control (Fig. \ref{Net2_2}) is more conservative compared with the optimal control needed for the real network (Fig. \ref{Net2_1}). Fig. \ref{fig:model_mismatch} further presents a case study to illustrate this phenomenon, where $\gamma_1=0.5,\gamma_2=0.3,\zeta=0.3$. Fig. \ref{fig:model_mismatch} corroborates that the optimal control designed under accurate modeling can successfully lead the system to a disease-free equilibrium. Furthermore, the control designed for network in Fig. \ref{net1} is able to achieve the disease-free objective if applied to the network in Fig. \ref{net1_1} (shown in Fig. \ref{model_mismatch_2}), while not vice versa (shown in Fig. \ref{model_mismatch_1}). Therefore, when the system operator has uncertainties on the underlying network structure, it might be better to design a more conservative control strategy to achieve a disease-free objective, as inaccurate modeling is possible to yield an epidemic outbreak.

\section{Conclusion}\label{conslusion}
We have studied the optimal control of competing  epidemics spreading over complex networks. The competing mechanism between two strains of epidemics results in a non-coexistence phenomenon at the steady state. Furthermore, we have explicitly derived the conditions under which the network is stabilized at different equilibria with control. The optimal control computed via the designed iterative algorithm can effectively reduce the spreading of epidemics. At the disease-free equilibrium, the optimal control is independent of nodes' degree distribution as the optimal strategy can be fully determined by the sufficient statistics including the average degree and the second moment of the degree distribution. Furthermore, depending on the epidemic parameters, the network equilibrium can switch via the adopted control strategy. Once the epidemic network switches to the disease-free equilibrium under the optimal control, the applied effort does not increase though the unit cost of effort continues to decrease, and the optimal control effort at the associated switching point is called the fulfilling threshold. One possible direction for future work would be extending the framework to multi-strains scenario and considering heterogeneous types of epidemic interdependencies. Another direction is to explicitly consider the transient behavior of the epidemic spreading and develop time-varying optimal control strategies. It is also worth investigating the optimal curing design when the system operator or node does not have a perfect knowledge on the node's state in the network.
 
\bibliographystyle{IEEEtran}
\bibliography{IEEEabrv,references}

\end{document}